\documentclass[a4paper,fleqn,usenatbib]{mnras}

\usepackage{newtxtext,newtxmath}
\usepackage[T1]{fontenc}
\usepackage{ae,aecompl}
\usepackage{pdflscape}

\usepackage{graphicx}	% Including figure files
\usepackage{amsmath}	% Advanced maths commands
\usepackage{amssymb}	% Extra maths symbols

\usepackage[colorinlistoftodos,color=blue!20]{todonotes}

\newcommand\kms{km\,s$^{-1}$}

\newcommand\msol{{M}$_\odot$}

\newcommand\spy{\;\msol~yr$^{-1}$}
\title[ALMA view of CS and SiS around O-rich AGB stars]{An ALMA view of CS and SiS around oxygen-rich AGB stars}

\author[T. Danilovich et al.]{T. Danilovich$^{1}$\thanks{E-mail: taissa.danilovich@kuleuven.be}\thanks{Postdoctoral Fellow of the Fund for Scientific Research (FWO), Flanders, Belgium},
A. M. S. Richards$^{2}$,
A. I. Karakas$^{3}$,
M. Van de Sande$^{1}$,
L. Decin$^{1}$, and
\newauthor  F. De Ceuster$^{1,4}$
\\
$^{1}$ Department of Physics and Astronomy, Institute of Astronomy, KU Leuven, Celestijnenlaan 200D, 3001 Leuven, Belgium  \\
$^{2}$ JBCA, Department Physics and Astronomy, University of Manchester, 
              Manchester M13 9PL, UK\\
$^{3}$ Monash Centre for Astrophysics, School of Physics \& Astronomy, Monash University, Victoria 3800, Australia\\
$^{4}$ Department of Physics and Astronomy, University College London, Gower Place, London, WC1E 6BT, UK
}

\date{Accepted XXX. Received YYY; in original form 2018 October 24}

\pubyear{2018}

\begin{document}
\label{firstpage}
\pagerange{\pageref{firstpage}--\pageref{lastpage}}
\maketitle

% Abstract of the paper
\begin{abstract}
{We aim to determine the distributions of molecular SiS and CS in the circumstellar envelopes of oxygen-rich asymptotic giant branch stars and how these distributions differ between stars that lose mass at different rates.
In this study we analyse ALMA observations of SiS and CS emission lines for three oxygen-rich galactic AGB stars: IK~Tau, with a moderately high mass-loss rate of $5\times10^{-6}$\spy, and W~Hya and R~Dor with low mass loss rates of $\sim1\times10^{-7}$\spy. These molecules are usually more abundant in carbon stars but the high sensitivity of ALMA allows us to detect their faint emission in the low mass-loss rate AGB stars. The high spatial resolution of ALMA also allows us to precisely determine the spatial distribution of these molecules in the circumstellar envelopes. 
We run radiative transfer models to calculate the molecular abundances and abundance distributions for each star. We find a spread of peak SiS abundances with $\sim10^{-8}$ for R~Dor, $\sim10^{-7}$ for W~Hya, and $\sim3\times10^{-6}$ for IK~Tau relative to H$_2$. We find lower peak CS abundances of $\sim7\times10^{-9}$ for R~Dor, $\sim7\times10^{-8}$ for W~Hya and $\sim4\times10^{-7}$ for IK~Tau, with some stratifications in the abundance distributions. For IK~Tau we also calculate abundances for the detected isotopologues: C$^{34}$S, $^{29}$SiS, $^{30}$SiS, Si$^{33}$S, Si$^{34}$S, $^{29}$Si$^{34}$S, and $^{30}$Si$^{34}$S.  Overall the isotopic ratios we derive for IK~Tau suggest a lower metallicity than solar. %{ What is the point of the paper?}
%It should be a single paragraph not more than 250 words (200 words for Letters).
}
\end{abstract}

% Select between one and six entries from the list of approved keywords.
% Don't make up new ones.

%%%%%%%%%%%%%%%%%%%%%%%%%%%%%%%%%%%%%%%%%%%%%%%%%%

%%%%%%%%%%%%%%%%% BODY OF PAPER %%%%%%%%%%%%%%%%%%

\section{Introduction}

The asymptotic giant branch (AGB) is a post-main sequence stage in the stellar evolution of low- to intermediate-mass stars with masses in the range 0.8--8~\msol{} \citep[][and references therein]{Herwig2005,Hofner2018}. The AGB phase is characterised by vigorous mass loss, with ejected matter forming an expanding circumstellar envelope (CSE). Molecules and dust grains form in the circumstellar envelope and are eventually returned to the interstellar medium (ISM). In this way, AGB stars are a significant source of chemical enrichment of the ISM and contribute to the chemical evolution of galaxies \citep{Romano2010,Kobayashi2011,Prantzos2018}.

The chemical composition of the CSE depends on the chemical type of the AGB star, which is often assigned through optical spectral classification. The primary spectral classifications are based on the C/O ratio which in turn plays a key role in determining the chemical characteristics of the CSE. Oxygen-rich AGB stars, which have C/O < 1, are the most common. Stars arrive on the AGB oxygen rich and over time carbon (and other nucleosynthesis products) may be dredged up from the interior and lead to an increase in {the surface} C/O and hence a shift in the chemical properties of the CSE. Some AGB stars may become carbon-rich over time with C/O > 1 (and are believed to pass through the transitionary phase of S-type stars with C/O $\sim1$),  
{while others are massive enough to experience hot bottom burning at the base of the convective envelope; this process provokes the destruction of the surface carbon, leaving the stars oxygen-rich} {\citep{Renzini1981,Bloecker1991,Boothroyd1991,Boothroyd1992,Lattanzio1992}. For a review on AGB evolution and nucleosynthesis including HBB models we refer to \cite{Karakas2014}. }

Sulphur is not synthesised in AGB stars nor in their main sequence progenitors. This allows us to constrain the study of sulphur-bearing molecules in AGB CSEs since the total abundance of sulphur can be estimated (usually based on solar abundance when considering nearby stars) and does not change significantly over time, since it is not thought to be depleted onto dust \citep{Waelkens1991,Reyniers2007}. Sulphur is also the tenth most abundant element in the universe, meaning that when a significant portion is locked up in particular molecules, they are relatively easy to detect in observations. The most common sulphur-bearing molecules found in AGB CSEs are CS, SiS, SO, SO$_2$ and H$_2$S. These have been most commonly studied using spatially unresolved observations of rotational transition lines, for example by \cite{Danilovich2017a} for H$_2$S.

In a study focussing on CS and SiS, \cite{Danilovich2018} surveyed a sample of AGB stars covering a range of mass-loss rates and chemical types. They detected CS in all surveyed carbon stars, some of the S-type stars and the highest mass-loss rate M-type stars. SiS was only detected in the highest mass-loss rate sources across chemical types. The sensitivity of their observations and subsequently calculated upper limits were insufficient to determine whether the high mass-loss rate M-type AGB stars genuinely have higher abundances of CS and SiS than the low mass-loss rate M-type AGB stars. More sensitive observations were required to conclusively make that determination.
Such sensitive observations are possible with telescopes such as the Atacama Large Millimetre/sub-millimetre Array (ALMA), which, with its high spatial resolution, also allows us to accurately map the distribution of circumstellar emission. For example, \cite{Brunner2018} recently presented ALMA maps and radiative transfer models of several molecular species towards the S-type AGB star W~Aql, including CS, SiS and $^{30}$SiS for which they were able to accurately determine the radial abundance distributions. %Their results were in general agreement with the other stars studied by \cite{Danilovich2018}.

{ To check the abundances of CS and SiS in lower mass-loss rate M-type stars, we must look to the available sensitive ALMA observations of such stars. Two of the closest AGB stars are R Dor and W Hya, which are both oxygen-rich and have low mass-loss rates of $1.6\times 10^{-7}$\spy{} for R~Dor \citep{Maercker2016} and $1\times 10^{-7}$\spy{} for W~Hya \citep{Khouri2014}. The most significant difference between W~Hya and R~Dor is that the former is a Mira variable while the latter is a semi-regular variable of type B (SRb). Both have been subject to APEX observations which did not detect CS or SiS (see \cite{De-Beck2018} for R~Dor and \cite{Danilovich2018} for W~Hya). Their proximity makes them ideal to search for weak CS and SiS emission in ALMA observations. Detected emission lines from these molecules will allow us to determine abundances, while non-detections will allow us to place more stringent upper limits on their abundances. }

{ Both R~Dor and W~Hya have recently been observed by ALMA, as detailed in Sect. \ref{obs}. Also observed was IK~Tau, a higher mass-loss rate AGB star losing $5.0\times 10^{-6}$\spy{} \citep{Maercker2016}. CS and SiS emission was detected towards IK~Tau by APEX and analysed in \cite{Danilovich2018}. The spatially-resolved ALMA observations of these molecules will allow us to precisely determine their radial abundance distributions and hence compare these with the modelling results of \cite{Danilovich2018}, which are based on single-dish observations. This will enable us to check the reliability of the empirical formulae found by \cite{Danilovich2018} for finding CS and SiS $e$-folding radii, and to compare the similarly determined abundance distributions with those for the low mass-loss rate stars. }

%A recently published spectral scan by \cite{Decin2018} of two oxygen-rich AGB stars obtained spatially resolved and sensitive observations of 197 lines towards R~Dor and 168 lines towards IK~Tau from 15 molecular species (with some lines still unidentified). R~Dor is a low mass-loss rate AGB star with a mass loss rate of $1.6\times 10^{-7}$\spy{} and IK~Tau has a higher mass-loss rate of $5.0\times 10^{-6}$\spy{} \citep{Maercker2016}. Several transition lines from sulphur-bearing molecules were detected with ALMA towards both stars, most notably SO and SO$_2$, for which several lines were covered by the spectral scan. In the case of IK~Tau, there were also clear detections of SiS, CS, and NS. 
%W~Hya is another low mass-loss rate AGB star with a mass loss rate of $1\times 10^{-7}$\spy{} \citep{Khouri2014}. The most significant difference between W~Hya and R~Dor is that the former is a Mira variable while the latter is a semi-regular variable of type B (SRb). W~Hya was included in the study of \cite{Danilovich2018} where no CS or SiS were detected using APEX. However, with the higher sensitivity of ALMA both of these molecules are detected. 
%\todo[inline, color=red!20]{Insert info about W Hya here (see Leen's comments)}

Hence, in this study we will analyse the spatial distribution of CS and SiS detected with ALMA towards IK~Tau, W~Hya, and R~Dor. Using radiative transfer modelling, we will compare precisely determined abundance distributions with previously obtained (spatially unresolved) results.

\section{Observations}\label{obs}
\subsection{ALMA observations and data reduction\label{datared}}

Spectral scans of IK~Tau and R~Dor were taken with ALMA in the range 335--362 GHz (in Band 7) during August and September of 2015 (proposal 2013.1.00166.S, PI: L. Decin). The interferometer baseline lengths were in the range 40~m to 1.6~km, which allowed for imaging of structure up to angular scales of 2\arcsec{} and with angular resolution $\sim150$ mas. Image cubes were made for IK Tau with a channel resolution of 1.7~\kms{} with rms noise in line-free channels from 3 mJy around 340 GHz to 9 mJy at above 355 GHz. For R Dor the image cubes have a channel resolution of 0.8--0.9\kms{} with rms noise in the range 2.7--5.7 mJy.
The full survey is presented in \cite{Decin2018}, including a detailed discussion of the data reduction. Here we focus only on the CS and SiS emission observed as part of that survey.
%Towards both IK~Tau and R~Dor many lines from sulphur-bearing molecules were detected. Several lines of SO and SO$_2$ were detected towards both stars but CS, SiS, and NS were detected towards IK~Tau but not R~Dor. 
Various isotopologue lines of these molecules were also detected towards IK~Tau. 

The low mass-loss rate AGB star, W~Hya, was observed by ALMA with long baselines on 30 November and 3 and 5 December 2015 (proposal 2015.1.01446.S, PI: A. Takigawa). The  first results of these observations, focussing on AlO and $^{29}$SiO are presented in \cite{Takigawa2017}.
W Hya was observed on baselines from 17 m to 11 km, giving sensitivity to angular scales up to about 6~arcsec. Imaging using weighting for high resolution gave a synthesised beam of ($35\times28$) mas. The native spectral resolution was $\sim0.85$~km~s$^{-1}$.  We made image cubes for the whole data set using a higher weighting for short baselines for comparison with the R Dor and IK Tau data, giving a synthesised beam of about 0.1\arcsec{} (depending on frequency) and spectral resolution of 2 km s$^{-1}$. The rms noise in quiet channels also depends on frequency, being about 2.5 mJy, 2.1 mJy, and 2.2 mJy, in spectral windows 0, 1, and 2 respectively (around 330, 343, 345 GHz).
In all cases the images were made using the LSR (local standard of rest) velocity convention. The full width half maximum of the ALMA primary beam is about 15 arcsec in this frequency range and all of the results presented here are from the inner few arcsec where the reduction in sensitivity is negligible.

Where possible, we extracted azimuthally averaged radial profiles from the zeroth moment maps of the ALMA data. These are compared with our models in Sect. \ref{modelling}.
The uncertainties in the azimuthally averaged radial profiles come from different sources: fluctuations due to clumpiness or asymmetries in the distribution of the emission, both of which are most significant when the emission is strong, and, where the emission is weak, due to the uncertainty in the mean flux density being noise dominated. The error bars take these factors into account. See \cite{Decin2018} for further details regarding the observed radial profiles. The LSR velocity of each star, along with the right ascension and declination, is given in Table \ref{radec}.

\begin{table}
\caption{Stars included in our study.}
\label{radec}
\begin{center}
\begin{tabular}{cccc}
\hline\hline
Star & RA & Dec & $\upsilon_\mathrm{LSR}$ [\kms]\\
\hline
IK Tau & 03 53 28.9 & +11 24 21.9 & 34\phantom{.0}\\
W Hya & 13 49 02.0 &$-$28 22 03.5 & 40.5\\
R Dor & 04 36 45.6& $-$62 04 37.8 & \phantom{0}7.5 \\
\hline
\end{tabular}
\\{Notes:}  $\upsilon_\mathrm{LSR}$ is the stellar velocity relative to the local standard of rest.
\end{center}
\end{table}%

\begin{table}
\caption{CS and SiS lines detected with ALMA towards IK~Tau, W~Hya and R~Dor, including isotopologues.}\label{linelist}
\begin{center}
\begin{tabular}{cccrl}
\hline\hline
 Frequency & Molecule & Line & $E_\mathrm{up}$ & Star(s)\\
 $\mathrm{[GHz]}$ & & [$\arcsec$] & [K]\, &\\
\hline
342.8828 & $^{12}$C$^{32}$S & $7 \to 6$, $v = 0$ & 66 & all \\
\hline
337.3965 & $^{12}$C$^{34}$S & $7 \to 6$, $v = 0$ & 65 & IK Tau \\
\hline
341.4223 & $^{28}$Si$^{32}$S & $19 \to 18$, $v = 2$ & 2301 & IK Tau*\\
343.1010 & $^{28}$Si$^{32}$S & $19 \to 18$, $v = 1$ & 1237 & IK Tau \\
344.7795 & $^{28}$Si$^{32}$S & $19 \to 18$, $v = 0$ & 166 & all \\
%357.6060 & $^{28}$Si$^{32}$S & $20 \to 19$, $v = 3$ &  & IK Tau* \\
359.3732 & $^{28}$Si$^{32}$S & $20 \to 19$, $v = 2$ & 2318 & IK Tau \\
361.1403 & $^{28}$Si$^{32}$S & $20 \to 19$, $v = 1$ & 1254 & IK Tau \\
\hline
336.8150 & $^{29}$Si$^{32}$S & $19 \to 18$, $v = 1$ & 1224 & IK Tau* \\
338.4474 & $^{29}$Si$^{32}$S & $19 \to 18$, $v = 0$ & 163 & IK Tau \\
354.5241 & $^{29}$Si$^{32}$S & $20 \to 19$, $v = 1$ & 1241 & IK Tau* \\
356.2424 & $^{29}$Si$^{32}$S & $20 \to 19$, $v = 0$ & 180 & IK Tau* \\
\hline
348.3621 & $^{30}$Si$^{32}$S & $20 \to 19$, $v = 1$ & 1229 & IK Tau* \\
350.0356 & $^{30}$Si$^{32}$S & $20 \to 19$, $v = 0$ & 177 & IK Tau \\
\hline
339.9110 & $^{28}$Si$^{33}$S & $19 \to 18$, $v = 0$ & 163 & IK Tau \\
357.7829 & $^{28}$Si$^{33}$S & $20 \to 19$, $v = 0$ & 181 & IK Tau \\
\hline
335.3419 & $^{28}$Si$^{34}$S & $19 \to 18$, $v = 0$ & 161 & IK Tau \\
351.2792 & $^{28}$Si$^{34}$S & $20 \to 19$, $v = 1$ & 1235 & IK Tau* \\
352.9739 & $^{28}$Si$^{34}$S & $20 \to 19$, $v = 0$ & 178 & IK Tau \\
\hline
346.3086 & $^{29}$Si$^{34}$S & $20 \to 19$, $v = 0$ & 175 & IK Tau \\
\hline
%340.1013 & $^{30}$Si$^{34}$S & $20 \to 19$, $v = 0$ &  & IK Tau* \\
357.0883 & $^{30}$Si$^{34}$S & $21 \to 20$, $v = 0$ & 189 & IK Tau \\
\hline
\end{tabular}
\end{center}
(*) indicates lines participating in overlaps. 
\end{table}

\subsection{IK Tau}

\subsubsection{SiS}

\begin{figure*}
\begin{center}
\includegraphics[width=\textwidth]{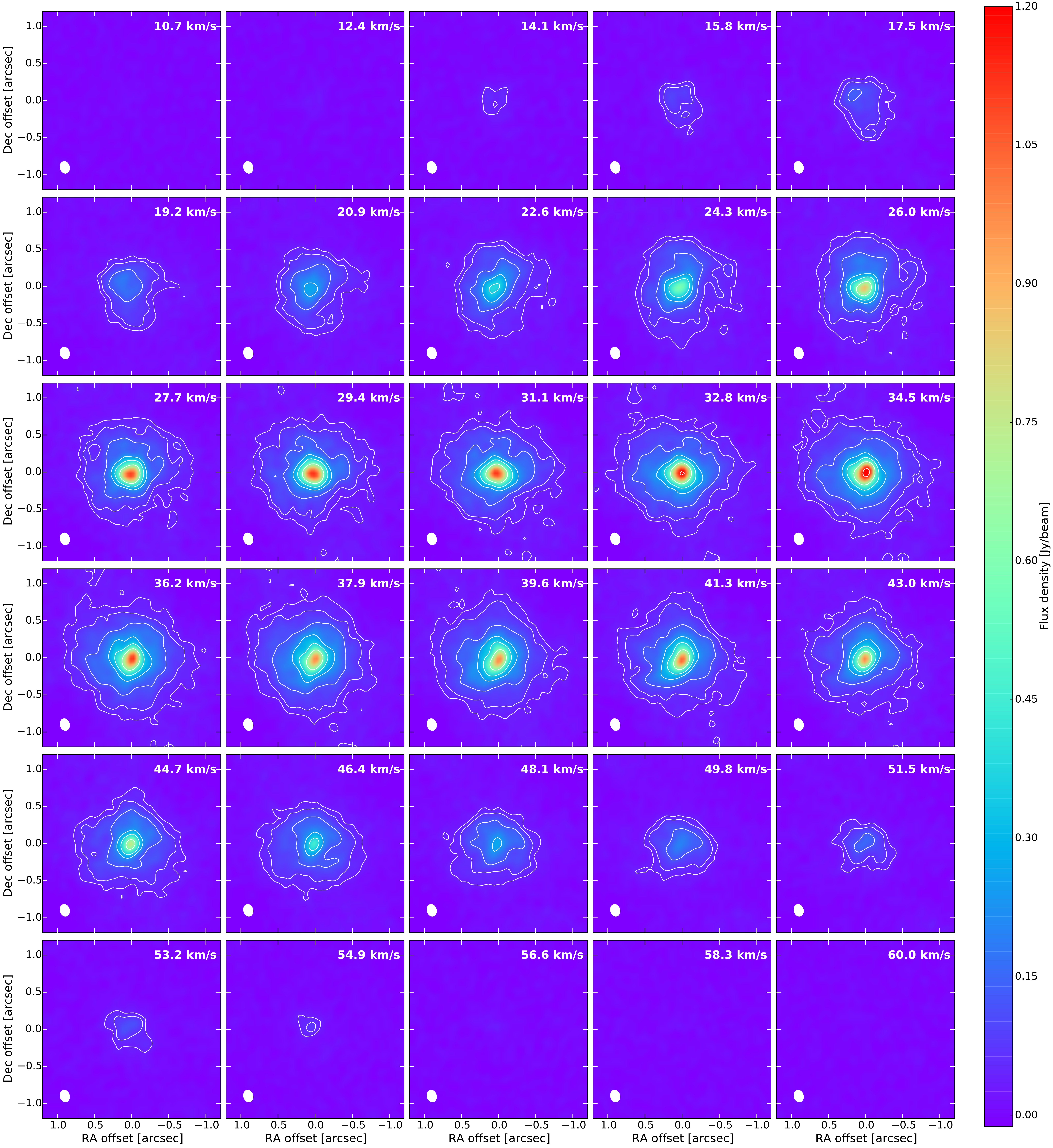}
\caption{The channel maps of the SiS ($19\to18$) transition at 344.7795 GHz for IK~Tau, which has $\upsilon_\mathrm{LSR} = 34$~\kms. The white contours indicate flux levels at 3, 5, 10, 20, 30, 50, and 100$\sigma$. The beam is indicated in white in the lower left corner of each channel.}
\label{sischanmaps}
\end{center}
\end{figure*}

\begin{figure}
\centering
\includegraphics[width=0.235\textwidth]{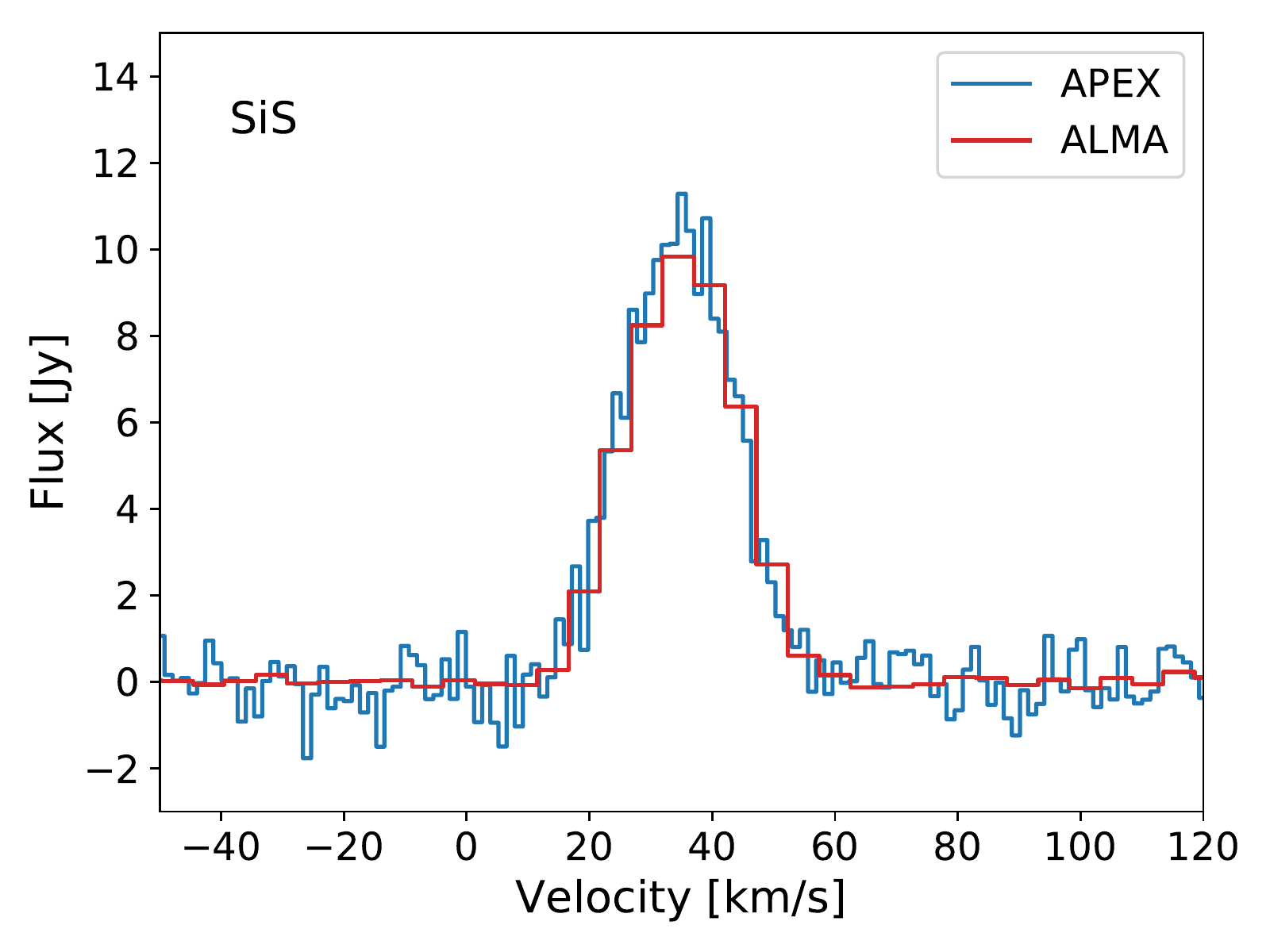}
\includegraphics[width=0.235\textwidth]{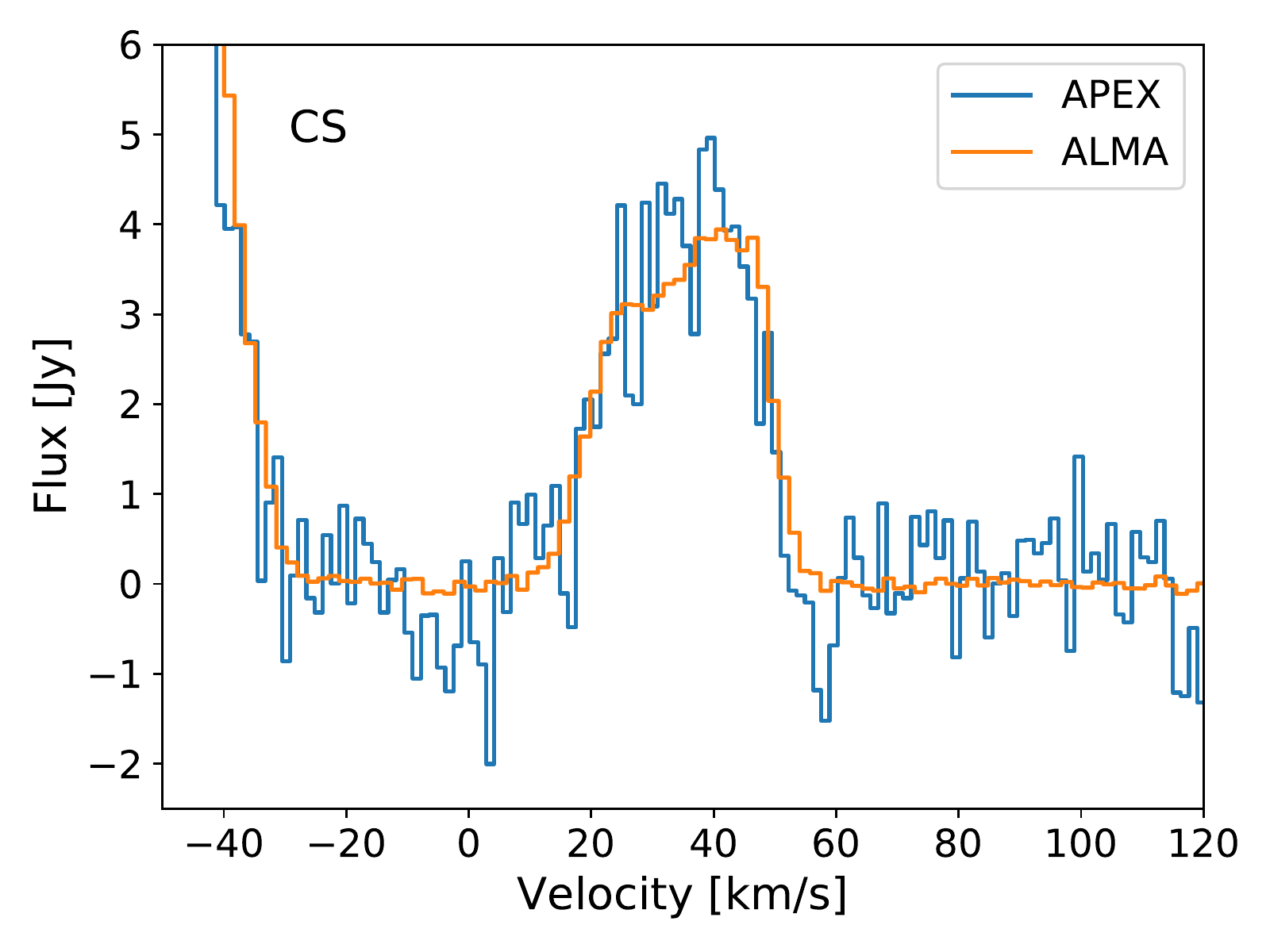}
\caption{A comparison of the APEX and ALMA observations of CS and SiS towards IK~Tau. \textit{Left:} SiS ($19\to18$); \textit{Right:} CS ($7\to6$). The red line for SiS indicates an ALMA spectrum extracted using a 5\arcsec radius and the orange line for CS indicates an ALMA spectrum extracted with an 800~mas radius aperture.}% (The adjacent blue line is \up{29}SiO ($8\to7$).)}
\label{iktaulostflux}
\end{figure}

The SiS ($19\to18$) transition at 344.7795 GHz was clearly detected towards IK~Tau with ALMA. The channel maps for this SiS transition are shown in Fig. \ref{sischanmaps}. %, especially in the weaker emission. 
%The SiS emission has a higher flux density than the CS emission show in Fig. \ref{cschanmaps}
In Fig. \ref{iktaulostflux} we compare the line spectrum extracted from the ALMA observation, using a 5\arcsec radius extraction aperture, with the spectrum observed for the same line by APEX. From this we can see that no flux has been resolved out in the ALMA observations.

Additionally, several isotopologue lines of SiS were detected. As well as the main isotopologue of $^{28}$Si$^{32}$S, ALMA detected lines from $^{29}$Si$^{32}$S, $^{30}$Si$^{32}$S, $^{28}$Si$^{33}$S, $^{28}$Si$^{34}$S and $^{29}$Si$^{34}$S. For the more abundant isotopologues, vibrationally excited lines were also seen, including two $v=2$ lines of $^{28}$Si$^{32}$S. All of these detected lines and their frequencies are listed in Table \ref{linelist}. Although a line of $v=3$ $^{28}$Si$^{32}$S and an additional line of $^{30}$Si$^{34}$S were noted in \cite{Decin2018}, they are too heavily blended with brighter overlapping lines for useful analysis and hence we exclude them here.

%\begin{figure}
%\begin{center}
%\includegraphics[width=0.5\textwidth]{IKTau-SiS(19-18,v=0)-Mom0-gist_ncar.pdf}
%\caption{The zeroth moment map of the SiS ($19\to18$) transition at 344.7795 GHz. The black contours show 99\% and 1\% levels of the IK~Tau continuum emission. The beam is indicated in white in the lower left corner.}
%\label{sismom0}
%\end{center}
%\end{figure}

%
%\begin{figure}
%\centering
%\includegraphics[width=0.5\textwidth]{IKTau-SiS-ALMA+APEX.pdf}
%\caption{A comparison of the APEX and ALMA observations of the IK~Tau SiS ($19\to18$) line. Note that the discrepancy between the APEX and ALMA line profiles are most likely due to calibration uncertainties with APEX.}
%\label{sislostflux}
%\end{figure}

\subsubsection{CS}

\begin{figure*}
\begin{center}
\includegraphics[width=\textwidth]{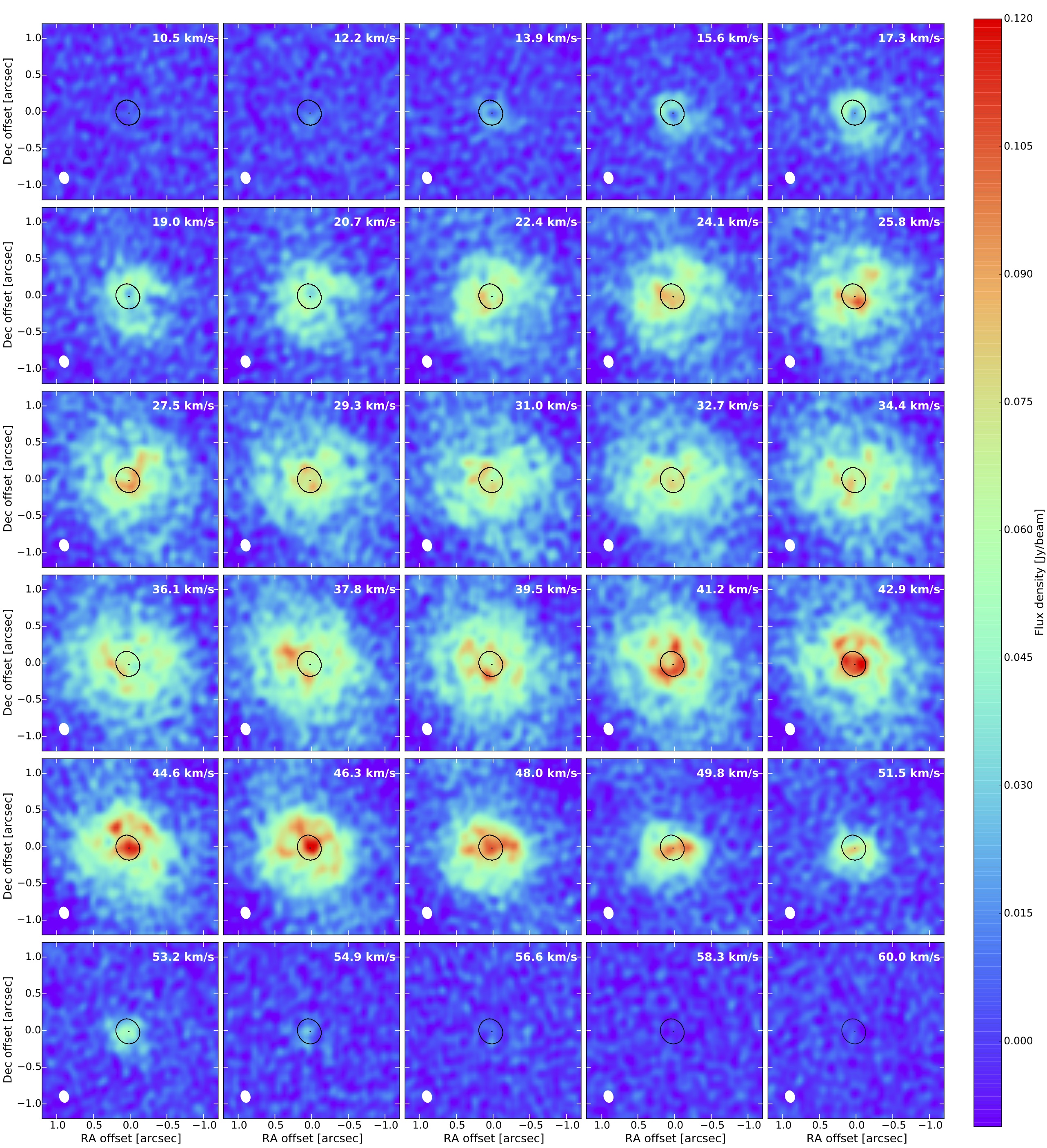}%should be pdf
\caption{The channel maps of the CS ($7\to6$) transition at 342.883 GHz for IK~Tau, which has $\upsilon_\mathrm{LSR} = 34$~\kms. The black contours show 90\% and 1\% levels of the IK~Tau continuum emission. The beam is indicated in white in the lower left corner of each channel.}
\label{cschanmaps}
\end{center}
\end{figure*}

The CS ($7\to6$) transition at 342.883 GHz was clearly detected towards IK~Tau with ALMA. The channel maps for this CS transition are shown in Fig. \ref{cschanmaps}. In Fig. \ref{iktaulostflux} we compare the line spectrum extracted from the ALMA observation with the spectrum observed for the same line by APEX. In this way we determine that no flux has been resolved out for this transition.

As can be seen in the channel maps, the peak in the CS emission is generally not centred on the continuum peak. The exception to this, in the channels around 44 to 46 \kms, is most likely due to the contribution of overlapping TiO$_2$ transitions at 342.8768 and 342.8618 GHz \citep{Decin2018}. In several of the central channels (around 41 to 46 \kms) there are brighter arcs partly resembling a ring of radius $\sim0.4\arcsec$ around the centre of the image. Together, the arcs and the lack of peak emission centred on the star suggest that CS is less abundant close to the star.
%We note that overall the CS emission appears a bit noisy, characterised by roughly beam-sized spots in the channel maps. This is most likely due the lower intensity of the CS emission (with peak intensity roughly 10\% that of the SiS emission plotted in Fig. \ref{sischanmaps}).

%A significant amount of noise can be seen in the CS emission, appearing as roughly beam-sized ``clumps'' in the channel maps in Fig. \ref{cschanmaps}. Nevertheless, it is evident in the channel maps that for most channels the CS emission does not peak at the continuum peak. There is also some asymmetry in the emission between the blue and the red sides of the $\upsilon_\mathrm{LSR} = 34$~\kms. This asymmetry is also clearly seen in the ALMA and APEX line profiles, in Fig. \ref{iktaulostflux}, and could be due to additional flux from blends due to TiO$_2$ lines at 342.8618 and 342.8768 GHz \citep{Decin2018}.

In addition to the main isotopologue of $^{12}$C$^{32}$S discussed above, we also weakly detected the ($7\to6$) transition of $^{12}$C$^{34}$S at 337.3965~GHz. Although it has a lower signal-to-noise ratio in the channel maps, it is visible in the zeroth moment map and in the spectrum.

%\begin{figure}
%\centering
%\includegraphics[width=0.5\textwidth]{IKTau-CS-ALMA+APEX.pdf}
%\caption{A comparison of the APEX and ALMA observations of the IK~Tau CS ($7\to6$) line. Note all of the flux has been recovered with ALMA.}% (The adjacent blue line is \up{29}SiO ($8\to7$).)}
%\label{cslostflux}
%\end{figure}

\subsection{W Hya}

\subsubsection{SiS}

\begin{figure*}
\begin{center}
\includegraphics[width=0.85\textwidth]{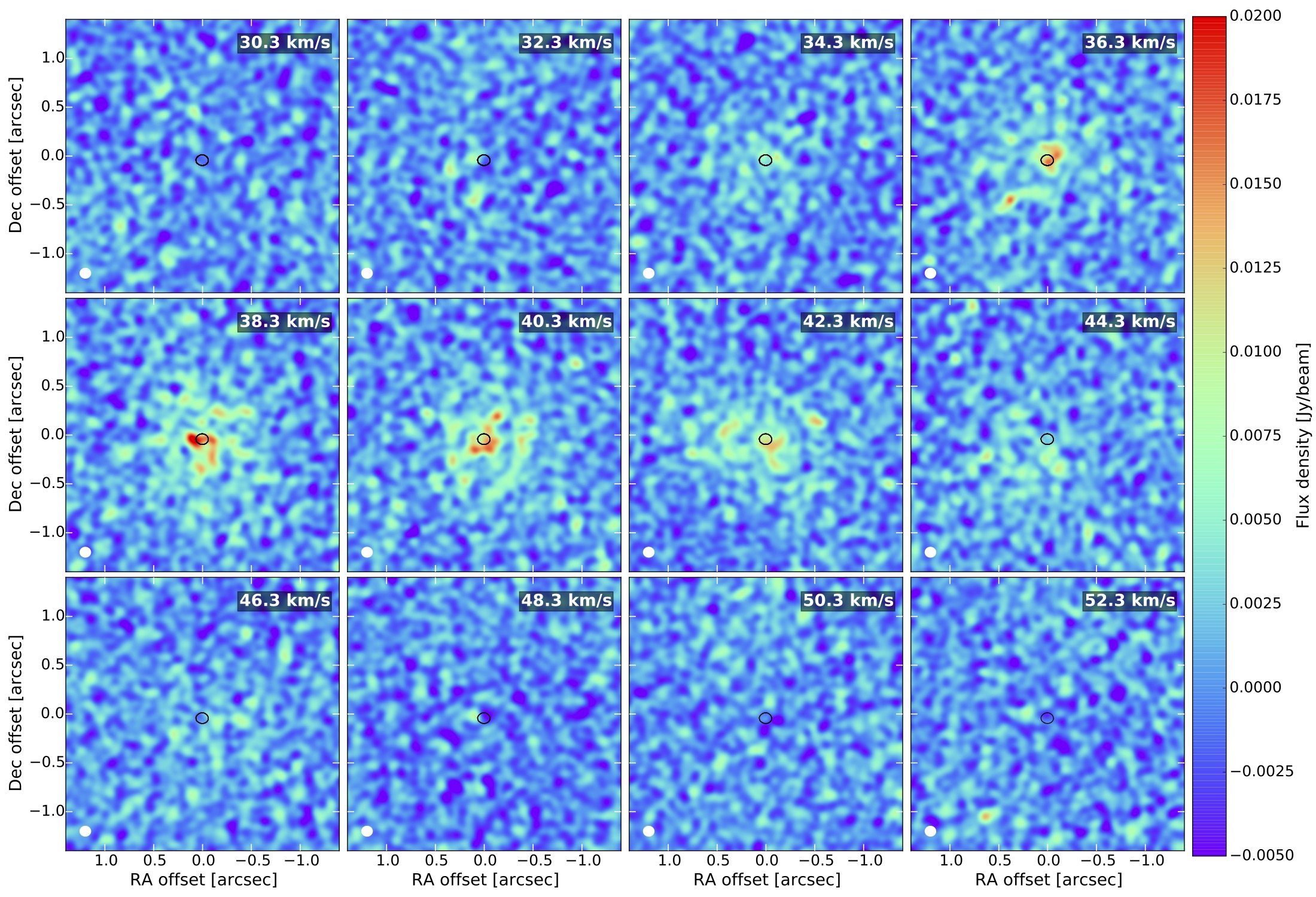} %should be pdf
\caption{The channel maps of the SiS ($19\to18$) transition at 344.7795 GHz for W~Hya, which has $\upsilon_\mathrm{LSR} = 40.5$~\kms. The black contours show 1\% of the peak W~Hya continuum emission. The beam is indicated in white in the lower left corner of each channel.}
\label{whyasischanmaps}
\end{center}
\end{figure*}

The SiS ($19\to18$) emission towards W~Hya is shown in channel maps in Fig. \ref{whyasischanmaps}. The emission is very weak and the signal-to-noise ratio is too low at this high spatial resolution to distinguish many features in the channel maps. The emission lines are clearly seen in the spectra extracted from the ALMA data. No other SiS lines were detected.
%\todo[inline]{Anita, do we have to worry about resolved out flux here?}

\subsubsection{CS}\label{whyacsobs}

\begin{figure*}
\begin{center}
\includegraphics[width=0.85\textwidth]{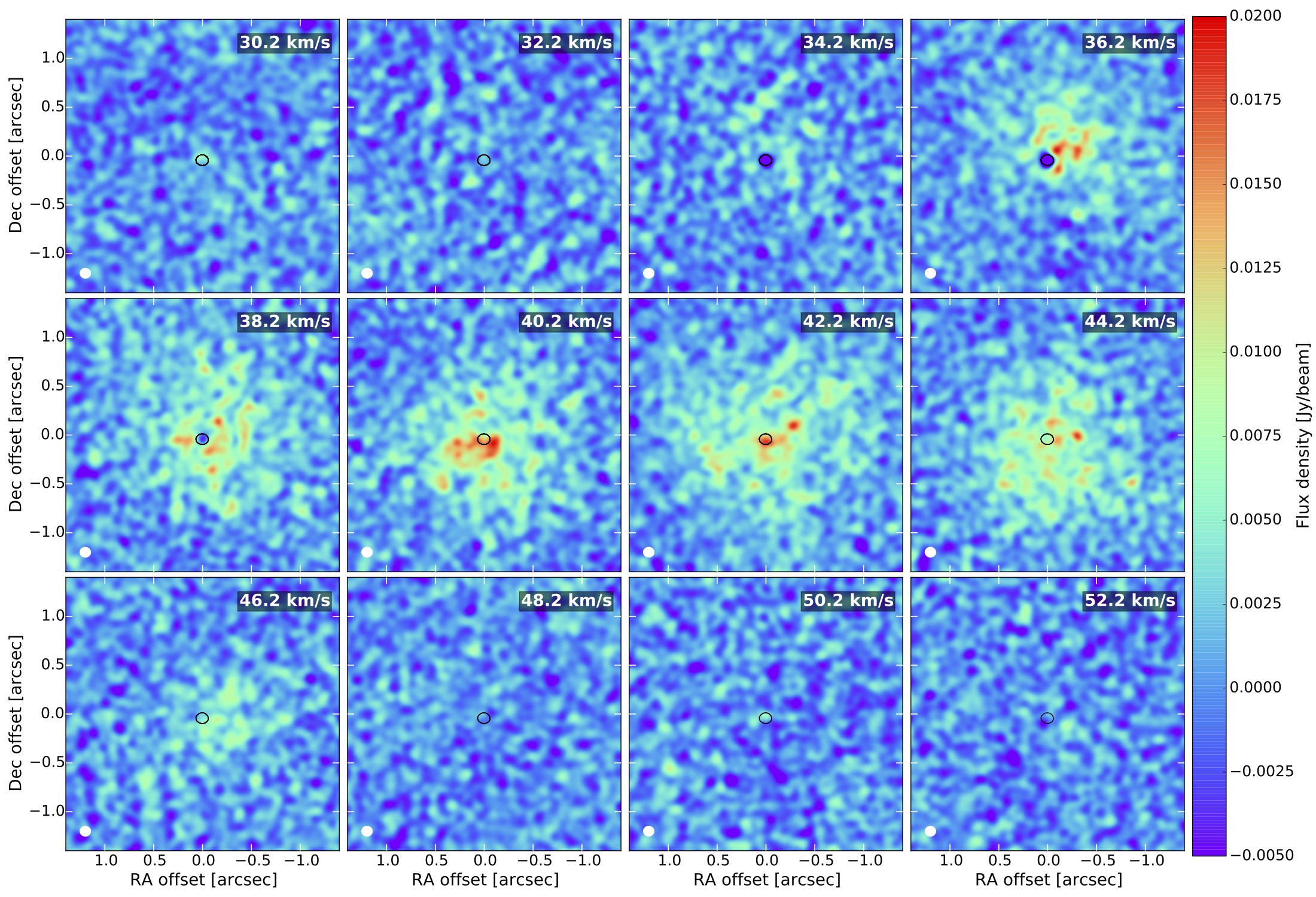}%should be pdf
\caption{The channel maps of the CS ($7\to6$) transition at 342.883 GHz for W~Hya, which has $\upsilon_\mathrm{LSR} = 40.5$~\kms. The black contours show 1\% of the peak W~Hya continuum emission. The beam is indicated in white in the lower left corner of each channel.}
\label{whyacschanmaps}
\end{center}
\end{figure*}

The channel maps for CS ($7\to6$) towards W~Hya are shown in Fig. \ref{whyacschanmaps}. The central absorption feature seen in the blue channels (those with velocities less than the LSR velocity $v_\mathrm{lsr} = 40.5$~\kms) is due to impact parameters along the light of sight to the star. It is seen here due to the high resolution of the observation such that there is a large ratio between the stellar angular diameter and the angular beam size. A similar phenomenon was seen for R~Dor by \cite{Decin2018}. In the spectra the absorption feature is clearly seen in the smallest extraction radius spectrum and the line is seen in emission for the larger radii extracted spectra.

\subsection{R Dor}

\subsubsection{SiS}

The ground state SiS ($19\to18$) line is not detected above the noise in the ALMA channel maps of R~Dor. However, if we extract a spectrum from the ALMA data in a region with a radius of 1\arcsec{} centred on the continuum peak, we do detect the SiS line. We use this detection and the tentative and non-detections in spectra extracted over smaller radii (300 mas and 75 mas, respectively) to constrain our SiS model for R~Dor in Sect. \ref{modelling}.

\subsubsection{CS}

The main CS line ($7\to6$) cannot be clearly seen in the channel maps of R~Dor.
% is visible in the 5 channel smoothed maps
%If we average the channels to obtain a lower velocity resolution of $\sim2.2$~\kms, 
However, if we extract a spectrum from the ALMA data in a region with a radius of 1\arcsec{} centred on the continuum peak, we do detect the CS line. A tentative detection is also seen in the spectrum from a 300~mas radius region centred on the continuum. For the spectrum extracted for a 75~mas region we do not detect emission but do tentatively detect an absorption feature that corresponds to the blue absorption discussed for other molecules in \cite{Decin2018}. We use this detection and the tentative detections to constrain our CS model for R~Dor in Sect. \ref{modelling}.

\section{Radiative transfer modelling}\label{modelling}

\subsection{Modelling procedure}\label{modproc}

We use a one-dimensional, spherically symmetric model to approximate the molecular emission of CS and SiS. Although this precludes the inclusion of asymmetric structures, we are still able to approximate the overall shape of the emission based on the average radial profiles calculated from the zeroth moment maps, and hence take into account various radial abundances and mean densities. 

The modelling is done using an accelerated lambda iteration method (ALI) which has been used to model a variety of molecular emission in the past, including both CS and SiS \citep{Brunner2018,Danilovich2018}. Where possible we have included previously observed single-dish observations to add constraints to our models \citep[see][for details on these archival observations]{Danilovich2018}.
To compare our models with the azimuthally averaged ALMA radial profiles, we extracted synthetic radial profiles from our models and plotted them against the ALMA profiles. 
The models were adjusted until the best fit to the data was found. In the case where emission was too weak to extract a radial profile from the observations --- such as in the case of R~Dor and some of the weaker isotopologue lines towards IK~Tau --- the models were based solely on the extracted spectra.

\subsection{Input parameters}

The crucial stellar and circumstellar parameters\footnote{Note that $\upsilon_\mathrm{exp}$, obtained from fitting CO spectra from single-dish observations, is the input parameter required to reproduce bulk of the line emission. It is not directly related to the higher velocities observed in the wings of various R~Dor and IK~Tau emission lines by \cite{Decin2018}.} that go into our radiative transfer models are listed in Table \ref{stellarprop}. All parameters are taken from \cite{Danilovich2017} and the references therein. {We note that the effective temperatures reported in Table 3, which are derived from SED fitting, are significantly cooler than stellar evolution models predict. However, a full examination of this issue is beyond the scope of the present paper.}
% These properties, most notably the mass-loss rates, were taken from \cite{Maercker2016} and \cite{Danilovich2016}.

\begin{table}
\caption{Stellar properties and input from CO models.}\label{stellarprop}
\begin{center}
\begin{tabular}{lccc}
\hline\hline
 Parameters       &       IK~Tau  &       R~Dor   &        W~Hya   \\
\hline                                                                                  
Luminosity [L$_\odot$]  &       7700    &       6500    &       5400  \\
Distance [pc]   &       265     &       59      &       78       \\
%$\upsilon_\mathrm{LSR}$ [\kms]        &       34      &       7       &       40.5   \\
$T_*$ [K]       &       2100    &       2400    &       2500       \\
%$R_\mathrm{in}$ [$10^{14}$ cm]  &       2.0     &       1.9     &       2.0     \\
%$\tau_{10}$     &       1.0     &       0.03    &       0.07       \\
$\dot{M}$ [$10^{-7}$~\spy]       &        $50$   &        $1.6$  &        $1$ \\
$\upsilon_\mathrm{exp}$ [\kms]      &       17.5    &       5.7     &       7.5     \\
%$\beta$         &       1.5     &       1.5     &       2.0     &       2.5     \\
\hline
\end{tabular}
\end{center}
{Notes:}  $T_*$ is the stellar temperature derived from SED fitting; $\dot{M}$ is the mass-loss rate; $\upsilon_\mathrm{exp}$ is the gas expansion velocity.
\end{table}%

\begin{figure}
\centering
\includegraphics[width=0.35\textwidth]{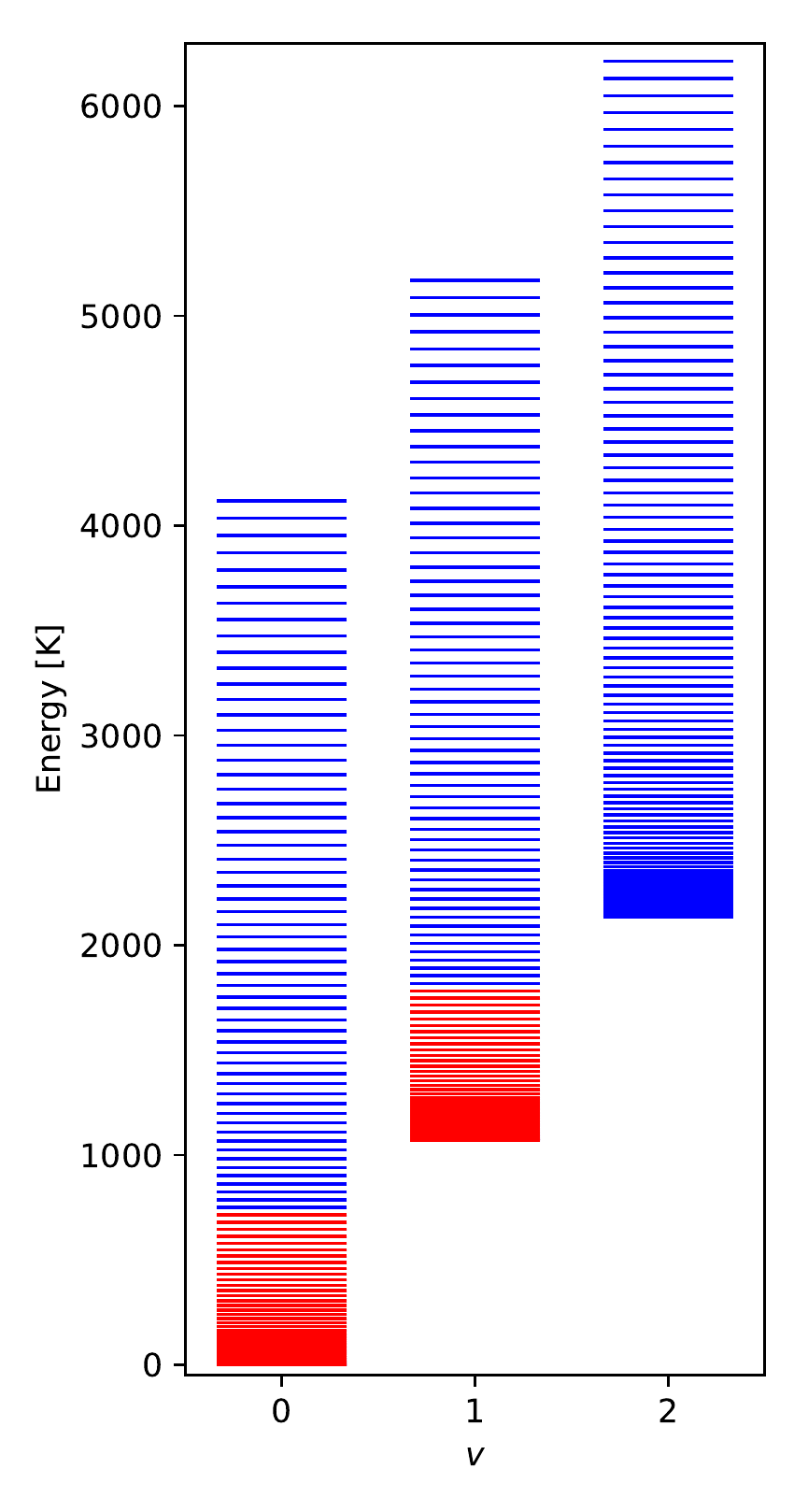}
\caption{$^{28}$Si$^{32}$S energy level diagram showing all the levels with $v=0,1,2$ and $0\leq J\leq99$ included in the expanded molecular description (see text for details). Levels highlighted in red are those with $v=0,1$ and $0\leq J\leq40$ included in the basic $^{28}$Si$^{32}$S molecular description.}
\label{siseld}
\end{figure}

We used molecular data for SiS that was previously implemented by \cite{Danilovich2018}: rotational energy levels from $J=0$ to $J=40$ in the ground and first excited vibrational states with parameters taken from the JPL spectroscopic database\footnote{\url{https://spec.jpl.nasa.gov}} \citep{Pickett1998} and the $v=0$ collisional rates for SiS-H$_2$ are scaled from the SiO-He rates of \citet{Dayou2006}. This set of energy levels is shown in red in Fig. \ref{siseld}. For the isotopologues, the same quantum numbered set of energy levels and identical collisional rates were used (since scaling for the difference in mass is negligible in these cases). The energy levels and radiative transition parameters for the isotopologues were taken from the Cologne Database for Molecular Spectroscopy \cite[CDMS\protect\footnote{\url{http://www.astro.uni-koeln.de/cdms}},][]{Muller2005,Endres2016} for $^{29}$SiS, $^{30}$SiS, and Si$^{34}$S and {from ExoMol\footnote{\url{http://exomol.com}} \citep{ExoMol,Upadhyay2018} for Si$^{33}$S, $^{29}$Si$^{34}$S, and $^{30}$Si$^{34}$S.} 

To facilitate the $v=2$ lines that were detected for the main isotopologue towards IK~Tau we created an expanded SiS data file from the molecular data available via CDMS \citep{Muller2007}. This expanded file included rotational energy levels from $J=0$ to $J=99$ and vibrational levels $v=0,1,2$, with the levels linked by all the relevant (ro)vibrational transitions. This expanded set of energy levels is shown in Fig. \ref{siseld} (in red plus blue). The same set of collisional rates, covering only levels $J\leq40$ in the ground vibrational state, were used since no vibrationally excited collisional rates are available. 
%SiS has a dipole moment of 1.735~D \citep{Hoeft1969,Murty1969,Muller2005}, so we expect it to be mainly radiatively excited and the impact of the collisional rates to be low. 
Since the $v=2$ lines were not detected for the less abundant isotopologues, we only use such an expanded molecular description for  $^{28}$Si$^{32}$S.

For CS  we also used the molecular data implemented by \cite{Danilovich2018}: rotational energy levels from $J=0$ to $J=40$ in the ground and first excited vibrational states were included with parameters taken from CDMS, and collisional rates based on those for CO-H$_2$ computed by \cite{Yang2010} were used. For the isotopologue model of C$^{34}$S we used energy levels and transitions for the same set of quantum states, also taken from CDMS, and the same set of collisional rates as for the main CS isotopologue.

\subsection{IK Tau model results}

\subsubsection{SiS}

\begin{figure}
\centering
\includegraphics[width=0.5\textwidth]{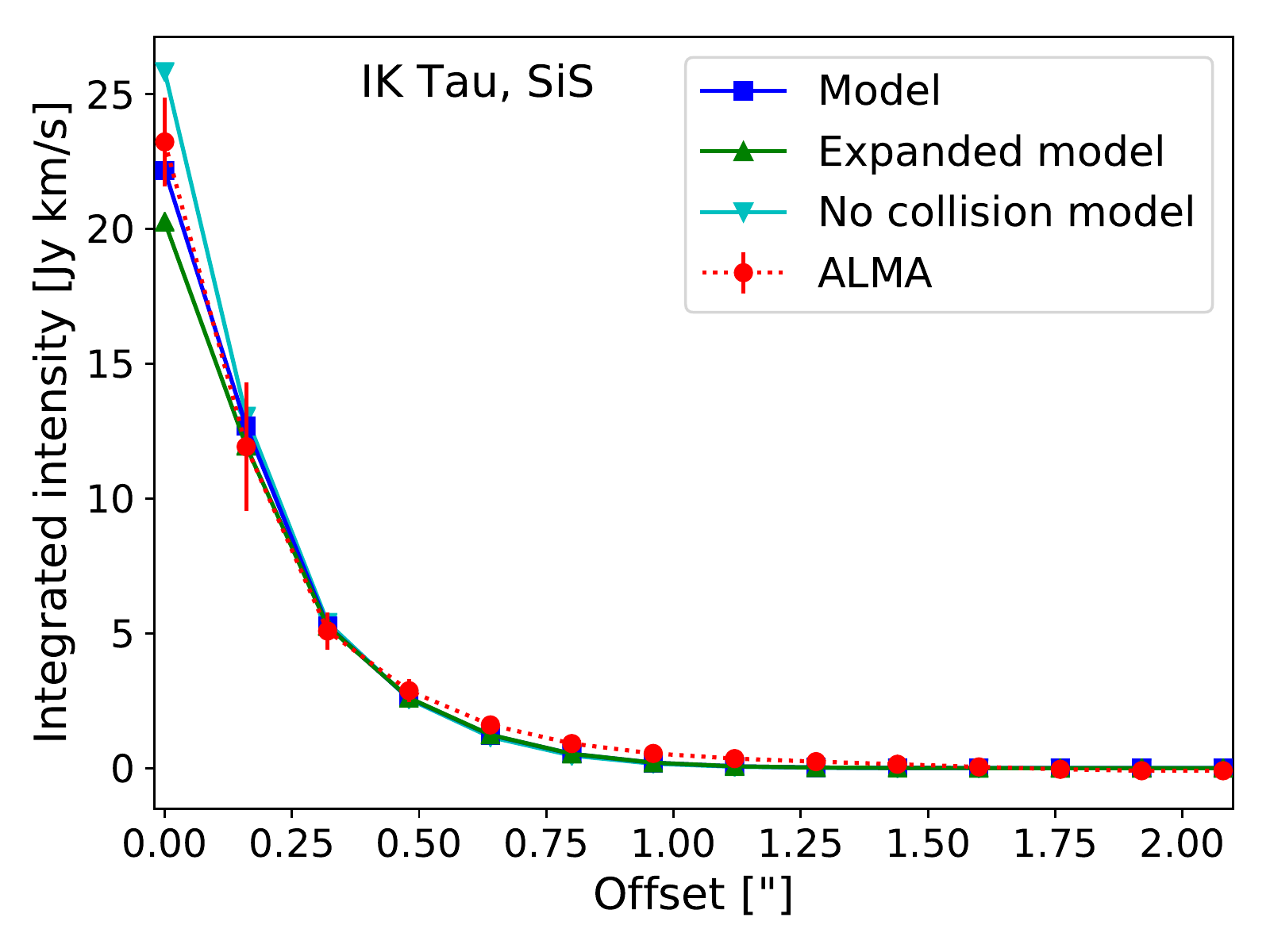}
\caption{The ALMA azimuthally averaged radial profile (red circles and dotted line) plotted with the best SiS ($19\to18$) model radial profile (blue squares and solid line) using the $v=0,1$ molecular description. The green line and triangles show the radial profile obtained for the same abundance profile when using the expanded ($v=0,1,2$) molecular description and the cyan line with inverted triangles shows the radial profile obtained for the expanded molecular description when neglecting collisions. The error bars shown on the ALMA data are as described in Sect. \ref{datared}.}
\label{sisrad}
\end{figure}

\begin{figure}
\includegraphics[width=0.5\textwidth]{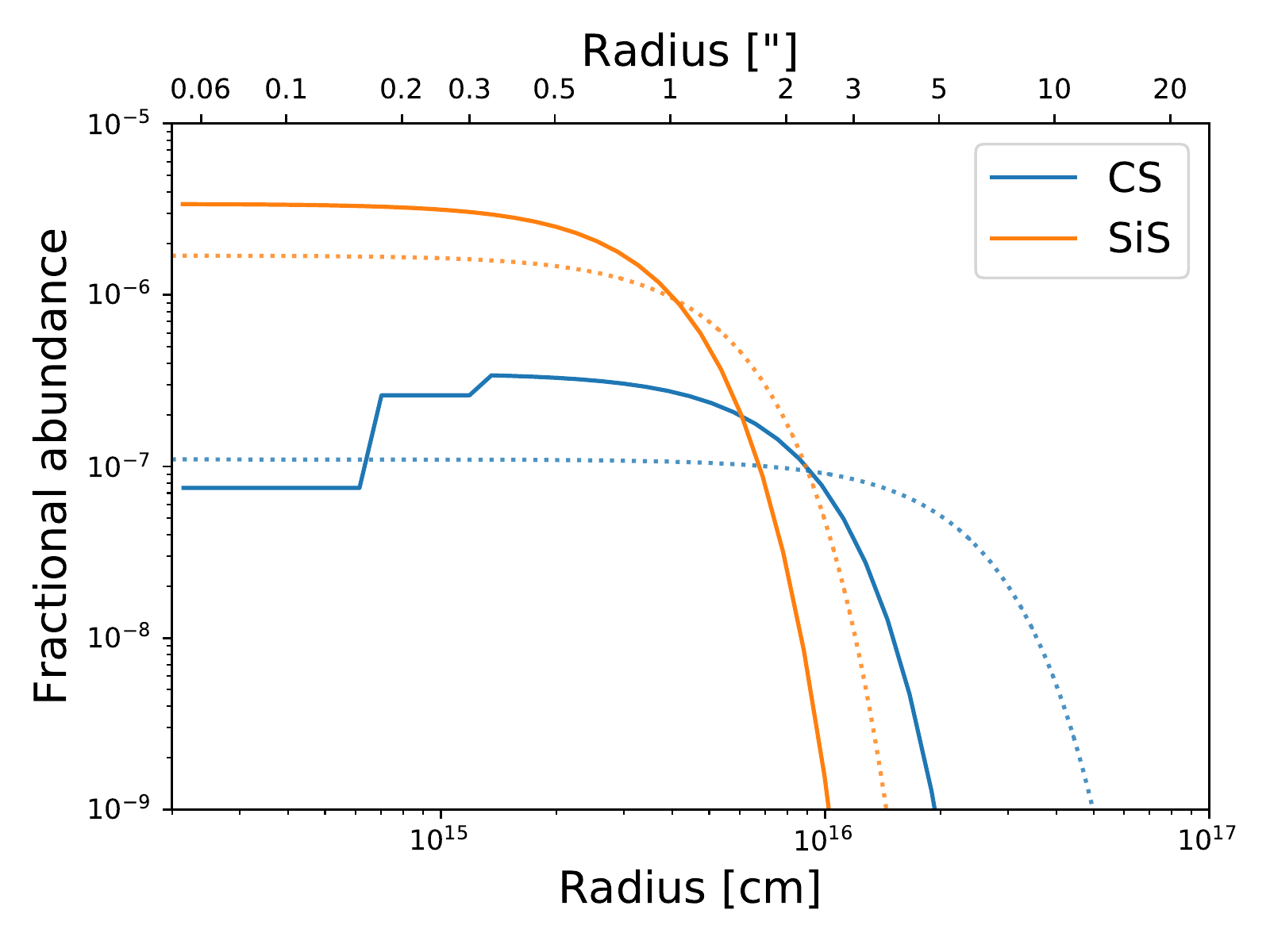}
\caption{The abundances derived from modelling the IK~Tau observations of CS and SiS. The solid lines are the abundance profiles derived from ALMA modelling in this study and the dotted lines are the abundance profiles derived from single-dish observations by \protect\cite{Danilovich2018}.}
\label{iktauabundance}
\end{figure}

For SiS towards IK~Tau, we were able to find a model that fits the data well using a Gaussian radial profile to describe the abundance stratification throughout the CSE. Since we also had access to several single-dish observations of SiS \citep[for details see][]{Danilovich2018}, we used these along with the ALMA spectra and radial profile (out to 0.75\arcsec) to constrain the model. The best fitting model found when only including the $v=0,1$ vibrational states has an inner fractional abundance relative to H$_2$ $f_0 = (3.4\pm0.5)\times 10^{-6}$ at an inner radius of $2\times10^{14}$~cm (about 4.3$R_*$) and an $e$-folding radius $R_e = (3.6\pm0.5)\times 10^{15}$~cm. The uncertainties calculated are based on a 90\% confidence interval. We experimented with smaller inner radii, but those models all significantly over-predicted the flux for the inner regions of the CSE. The final inner radius of $2\times10^{14}$~cm is that used by \cite{Danilovich2016} for the SO and SO$_2$ models of IK~Tau and is in agreement with the radius used by \cite{Maercker2016} for CO and H$_2$O.
The abundance profile of our final SiS model is plotted in Fig. \ref{iktauabundance} along with the abundance profile derived using only single-dish data by \cite{Danilovich2018}. The radial profile of this model is plotted in Fig. \ref{sisrad}.

\begin{figure*}
\begin{center}
\includegraphics[width=\textwidth]{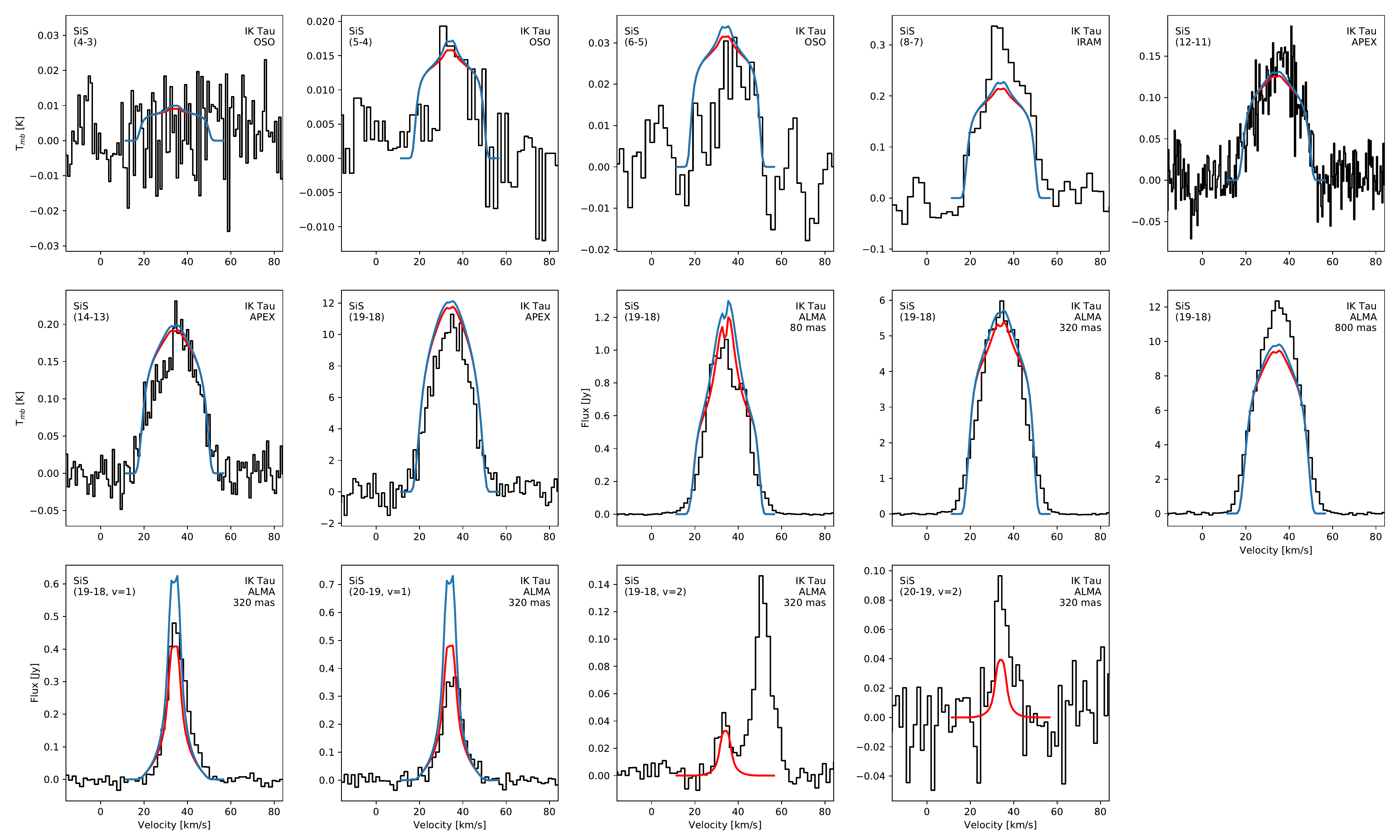}
\caption{IK~Tau observed SiS spectra (black histograms) with the corresponding lines from the best model (red lines) with molecular data up to $v=2$. For comparison, the model lines based on the molecular data up to $v=1$ are also included for comparison (blue lines). The angular sizes listed for the ALMA lines are the radii of the extraction apertures.}
\label{iktausis}
\end{center}
\end{figure*}

Running a model with the same abundance profile but with the expanded molecular description ($v=0,1,2$) gives a very similar result with the main differences being a reduction in intensity of the innermost radial profile point and better agreement with observations for the $v=1$ lines. This is expected since the availability of higher energy levels would have a more significant impact on the higher energy regions. The radial profile of the ($19\to18$) transition for this model is also plotted in Fig. \ref{sisrad} and the corresponding model lines for both molecular descriptions are plotted with single-dish observations and ALMA spectra in Fig. \ref{iktausis}.

We also tested the significance of the collisional rates by running another model with the same abundance profile but this time using the expanded molecular description ($v=0,1,2$) with all collisional rates set to zero --- effectively neglecting collisions. The result was again similar to the existing results with the only significant deviation being, again, in the innermost part of the radial profile. Very little change was seen in the outer parts of the radial profile and in the single dish observations. These changes make sense if we consider that collisions are expected to play a more significant role in the dense inner regions of the CSE.%, even if SiS is otherwise predominantly radiatively excited.

To determine the abundances of the SiS isotopologues, we used the same $e$-folding radius of $3.6\times 10^{15}$~cm that was found for $^{28}$Si$^{32}$S and varied the peak abundance until the best fit to the ALMA radial profile was found. Our models were equally weighted between the spectral lines extracted from the ALMA data and the radial profiles out to 0.4\arcsec, beyond which noise and instrumental effects made the data unreliable. For the two weakest isotopologues, $^{29}$Si$^{34}$S and $^{30}$Si$^{34}$S, only the spectra were used since the signal-to-noise ratio of the observations was too low to extract a radial profile.

The resultant isotopologue abundances are listed in Table \ref{isogaussresults} with various isotopologue ratios listed in Table \ref{sisratios}. Our convention has been to list the ratio such that the abundance of the isotopologue with the smaller mass number is divided by the abundance of the isotopologue with the larger mass number.
%The radial profiles of the SiS isotopologues are plotted in Fig. \ref{iktausisisoradprofs}.
There is good agreement between different isotopologues that trace the same isotopic ratios, suggesting that SiS isotopologues are good tracers of Si and S isotopic ratios. The only exception was $^{30}$Si$^{34}$S for which we found a higher abundance than expected. The ALMA spectra of $^{30}$Si$^{34}$S are both brighter and wider than those of $^{29}$Si$^{34}$S, which is unexpected since we otherwise obtain higher abundances of molecules with $^{29}$Si than $^{30}$Si (see Table \ref{isogaussresults}). The discrepancy is most likely due to the $^{30}$Si$^{34}$S ($21\to20$) line at 357.0883~GHz being contaminated by an unidentified blend. Excluding this line, the means of the isotopic ratios found are listed in Table \ref{isotopes} and compared with solar and literature values.

\begin{table}
\caption{IK Tau SiS isotopologue abundances for Gaussian models}\label{isogaussresults}
\begin{center}
\begin{tabular}{cc}
\hline\hline
 Molecule & $f_0$\\% &$^{29}$Si$^{32}$S& $^{30}$Si$^{32}$S& $^{28}$Si$^{33}$S &$^{28}$Si$^{34}$S\\
\hline
$^{28}$Si$^{32}$S & $(3.4\pm0.5)\times 10^{-6} $ \\%& 23 & 37 & 190 & 31\\
$^{29}$Si$^{32}$S & $(1.4\pm 0.2)\times 10^{-7}$  \\%& - &1.6 & - & -\\
$^{30}$Si$^{32}$S & $(8.4\pm1.0)\times 10^{-8}  $\\%& - & - & - &-\\
$^{28}$Si$^{33}$S & $(1.8\pm0.2)\times 10^{-8} $ \\%& -&-& - & 0.17\\
$^{28}$Si$^{34}$S & $(9.5\pm0.9)\times 10^{-8}$ \\%& -&- &- &-\\
$^{29}$Si$^{34}$S & $ (3.8\pm1.2)\times 10^{-9}$ \\%& $\goa30$ & - & - &$\goa22$\\
$^{30}$Si$^{34}$S & $ (1.1\pm0.3)\times 10^{-8}$ \\%& - &$\goa12$ & - & $\goa14$\\
\hline
\end{tabular}
\end{center}
\end{table}

\begin{table*}
\caption{IK Tau SiS isotopologue ratios, calculated as smaller mass numbers divided by larger mass numbers.}\label{sisratios}
\begin{center}
\begin{tabular}{cccccc}
\hline\hline
&\multicolumn{5}{c}{Isotopologue ratios}\\
 Molecule &$^{29}$Si$^{32}$S& $^{30}$Si$^{32}$S& $^{28}$Si$^{33}$S &$^{28}$Si$^{34}$S & $^{29}$Si$^{34}$S\\
\hline
$^{28}$Si$^{32}$S  & $24\pm5 $ & $40\pm8$ & $190\pm35$ & $36\pm6 $ & -\\
$^{29}$Si$^{32}$S   & - & $1.7\pm 0.3$ & - & - & $37\pm13$\\
%$^{30}$Si$^{32}$S & - & - & - &- & -\\
$^{28}$Si$^{33}$S  & -&-& - & $0.19 \pm 0.03$&-\\
$^{28}$Si$^{34}$S  & -&- &- &- & $25\pm 9$\\
$^{29}$Si$^{34}$S  & $37\pm 13$ & - & - &-&-\\
$^{30}$Si$^{34}$S  & - &$7.3\pm2$ & - & $8.6\pm2.5$ &$0.35\pm0.14$\\
\hline
\end{tabular}
\end{center}
\end{table*}	

\begin{table}
\caption{Mean isotope ratios derived from SiS towards IK Tau, with comparison to literature values for the Sun, R Dor, and W Hya}\label{isotopes}
\begin{center}
\begin{tabular}{ccccc}
\hline\hline
Ratio  & IK Tau& R Dor & W Hya & Solar$^{d}$ \\
\hline
$^{28}$Si/$^{29}$Si & $25 \pm 10$ & 12.6$^{a}$ & -& 19.7 \\
$^{28}$Si/$^{30}$Si & $40 \pm 8$ & 20.0$^{a}$ &-& 29.9 \\
$^{29}$Si/$^{30}$Si & $ 1.7\pm0.3$ & 1.58$^{a}$ & 0.99$^{c}$& 1.51  \\
$^{32}$S/$^{33}$S & $190\pm35$ & - & -& 125 \\
$^{32}$S/$^{34}$S & $37\pm14$ & 21.6$^{b}$ & -& 22 \\
$^{33}$S/$^{34}$S & $0.19\pm0.03$& - &-& 0.18 \\
\hline
\end{tabular}
\end{center}
References: ($^{a}$) \cite{De-Beck2018}; ($^{b}$) \cite{Danilovich2016}; ($^{c}$) \cite{Peng2013}; ($^{d}$) \cite{Asplund2009}
\end{table}

\subsubsection{CS}\label{iktaucsmodel}

\begin{figure}
\centering
\includegraphics[width=0.5\textwidth]{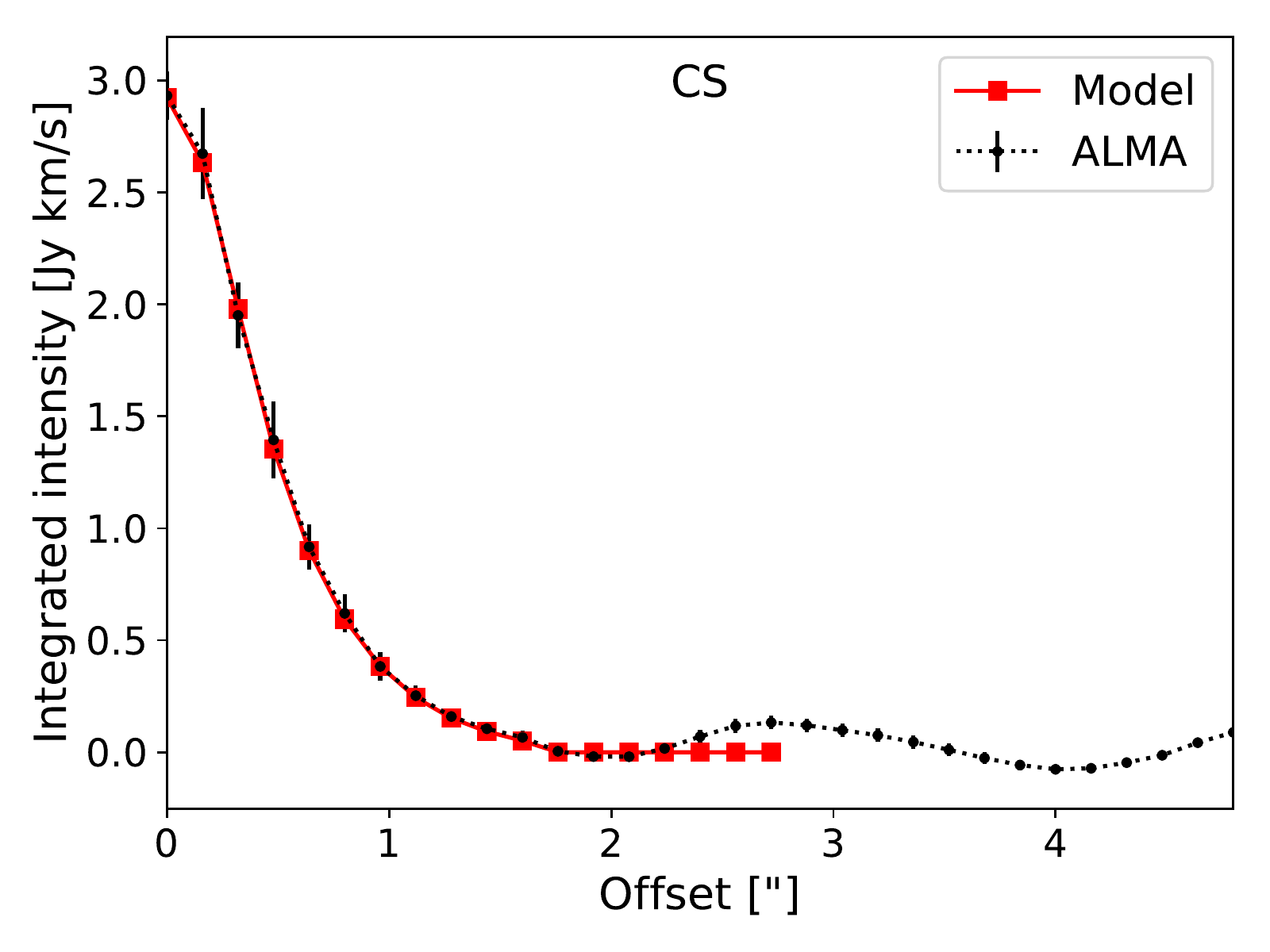}
\caption{The best CS model radial profile (red squares and solid line) plotted with the ALMA azimuthally averaged radial profile (black and dotted line) for the ($7\to6$) transition for IK~Tau. The error bars shown on the ALMA data are as described in Sect. \ref{datared}.\label{csrad}}
\end{figure}

We initially assumed a Gaussian abundance distribution for CS in the CSE of IK~Tau, starting from the same inner radius as for SiS. However, this strongly overpredicted the central brightness despite fitting the outer parts of the emission. Hence, we adjusted the radial abundance distribution so that there was a lower-abundance inner component with abundance $f_c$, out to a radius $R_c$, outside of which the original Gaussian abundance distribution was used. 
We found the model that best fits the ALMA radial profile had two stratified inner components, with $f_{c_1}=7.5\times10^{-8}$, $R_{c_1} = 7\times10^{14}$~cm, $f_{c_2}=2.6\times10^{-7}$, $R_{c_2} = 1.2\times10^{15}$~cm, and beyond $R_{c_2}$ a Gaussian component with $f_0 = 3.5\times10^{-7}$ and $R_e = 8\times 10^{15}$~cm. The radial profile from this model for the CS ($7\to6$) transition is shown with the ALMA azimuthally averaged profile in Fig. \ref{csrad}.
The derived abundance profile is shown in Fig. \ref{iktauabundance} where it is compared with the radial profile found by \cite{Danilovich2018} from fitting only single-dish CS data.

The observations of C$^{34}$S ($7\to6$) have a low signal-to-noise ratio. It also seems that some of the extended flux has been resolved out, based on some negative flux around 1\arcsec{} in the radial profile. Hence, we were unable to fit a model based purely on the observations. Instead we use a model with the same abundance structure as for C$^{32}$S and, taking the $^{32}$S/$^{34}$S ratio found from SiS, we find the CS observations to be in good agreement to the sulphur isotopic ratio found for SiS.

%Solar isotopologue ratios and our result/solar:
% 28/29 Si 19.7 lower than our result (127%)
% 28/30 Si 29.9 lower than our result (133%)
% 29/30 Si 1.51 a little lower than our result (113%)

% 32/33 S 125 lower than our result (152%)
% 32/34 S  22 lower than our result (168%)
% 33/34 S   0.18 in agreement with our results (105%)

\subsection{W Hya model results}

\subsubsection{SiS}

\begin{figure}
\centering
\includegraphics[width=0.5\textwidth]{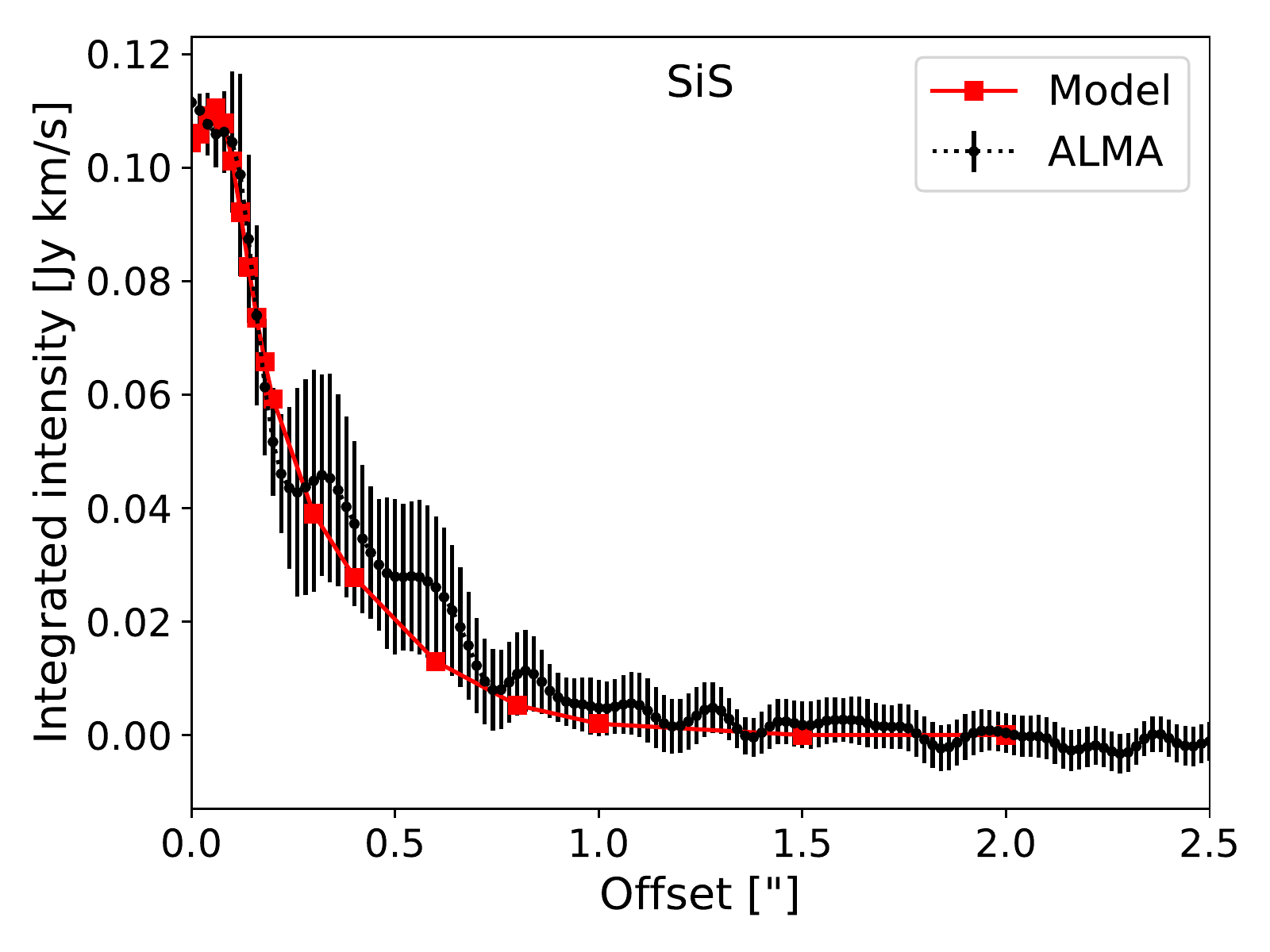}
\includegraphics[width=0.5\textwidth]{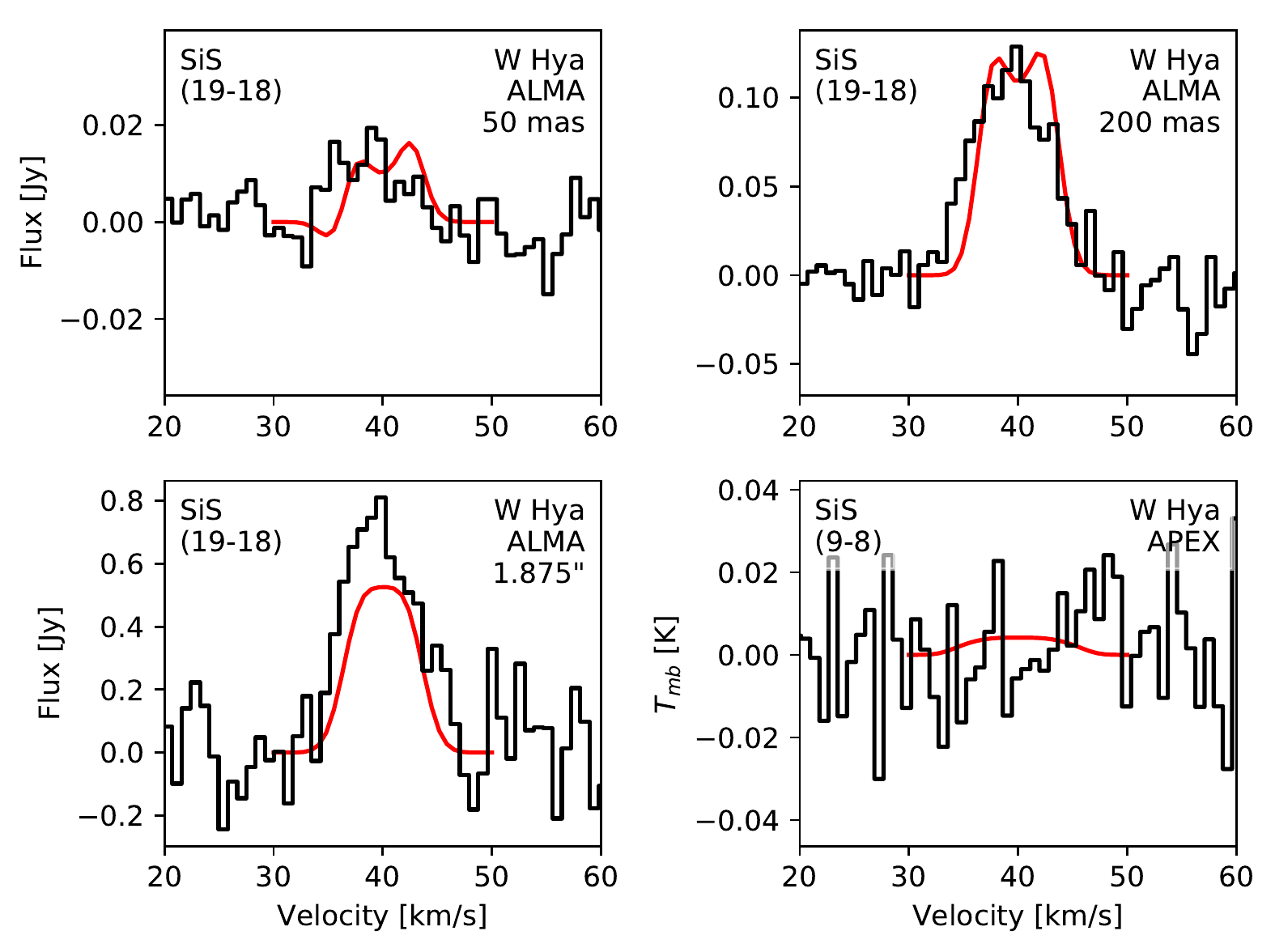}
\caption{\textit{Top:} The best SiS model radial profile (red squares and solid line) for the ($19\to18$) transition plotted with the ALMA azimuthally averaged radial profile (black points and dotted line) for W~Hya. The error bars shown on the ALMA data are as described in Sect. \ref{datared}. \textit{Bottom:} The observed spectra (black histograms) and model line profiles (red lines) for SiS towards W~Hya, including an observation from APEX, which serves as an upper limit. The sizes of the extraction radii for the ALMA spectra are listed in the top right corners. The emission line at $\sim$15~\kms{} in the ALMA spectrum is the 344.8079 GHz line of $^{34}$SO$_2$ ($13_{4,10} \to 13_{3,11}$). \label{whyasisrad}}
\end{figure}

\begin{figure}
\includegraphics[width=0.5\textwidth]{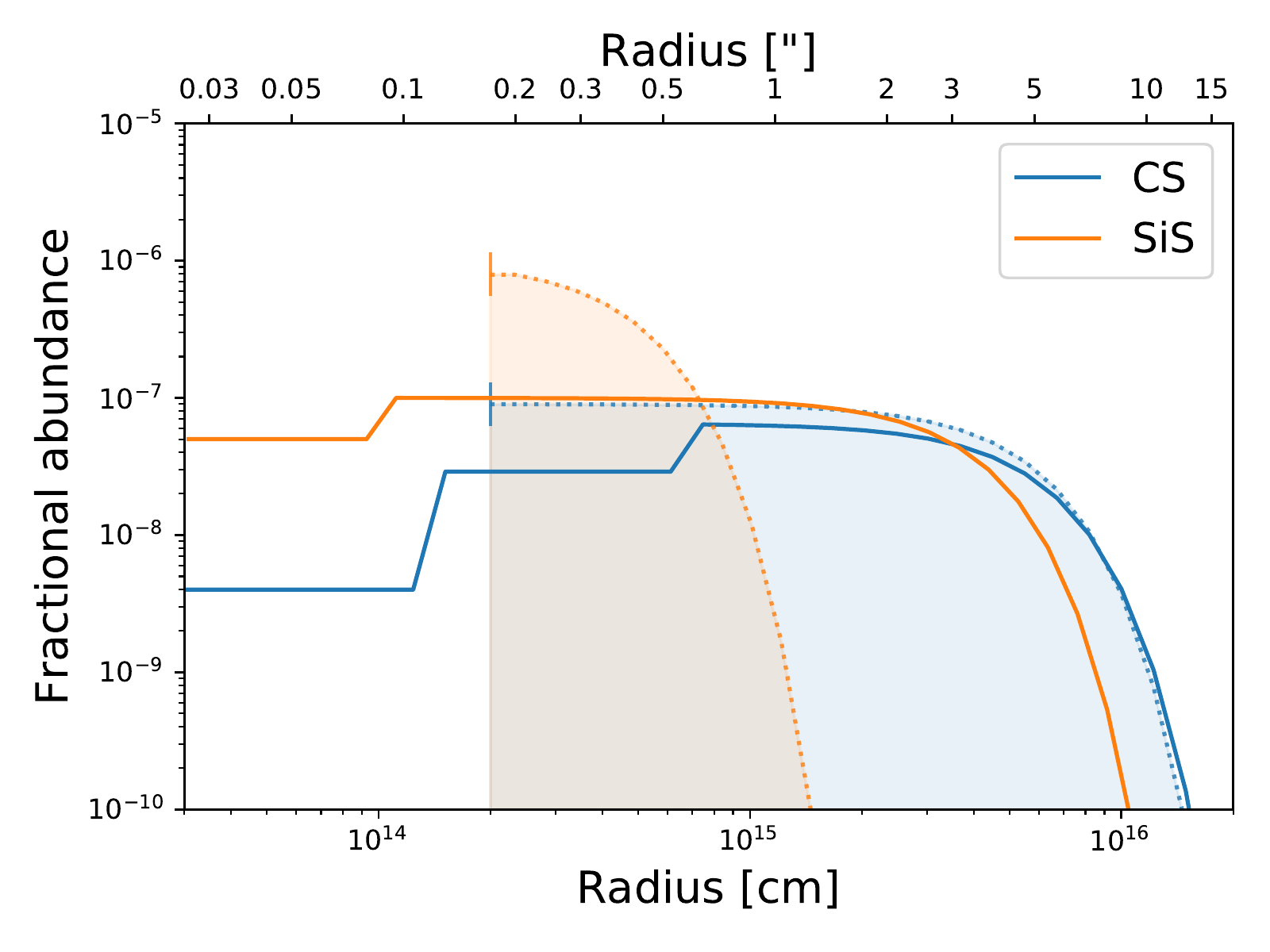}
\caption{The abundance distributions derived from modelling the W~Hya observations of CS and SiS (solid lines) and the upper limits for the same abundances derived by \protect\cite{Danilovich2018} from APEX non-detections (dotted lines with shaded regions), with the vertical lines indicating their model inner radii.}
\label{whyaabundance}
\end{figure}

For W~Hya, we used both the azimuthally averaged radial profile and the SiS spectra extracted from the ALMA data for different angular sizes. Two observations from \cite{Danilovich2018} of undetected lines were available: SiS ($9\to8$) and ($14\to13$). While both were included in our modelling (and did not contribute any additional constraints), we only plot one here. 
We found an adequate fit to the somewhat noisy radial profile using a model with a lower inner abundance followed by a Gaussian abundance distribution. Our W~Hya SiS model has a constant inner abundance of $f_c=5\times 10^{-8}$ starting from close to the stellar surface ($\sim R_*$) and running out to $R_c=1.1\times 10^{14}$~cm after which the Gaussian part of the distribution increases to $f_g=1\times 10^{-7}$ and has $R_e=4\times 10^{15}$~cm.
The radial profile for the ($19\to18$) transition from the best model and the ALMA observations is plotted in Fig. \ref{whyasisrad}, along with the modelled and observed spectral lines. The abundance profile of SiS for W~Hya is plotted in Fig. \ref{whyaabundance} and compared with the upper limit found by \cite{Danilovich2018} based on their non-detections.

\subsubsection{CS}

\begin{figure}
\centering
\includegraphics[width=0.5\textwidth]{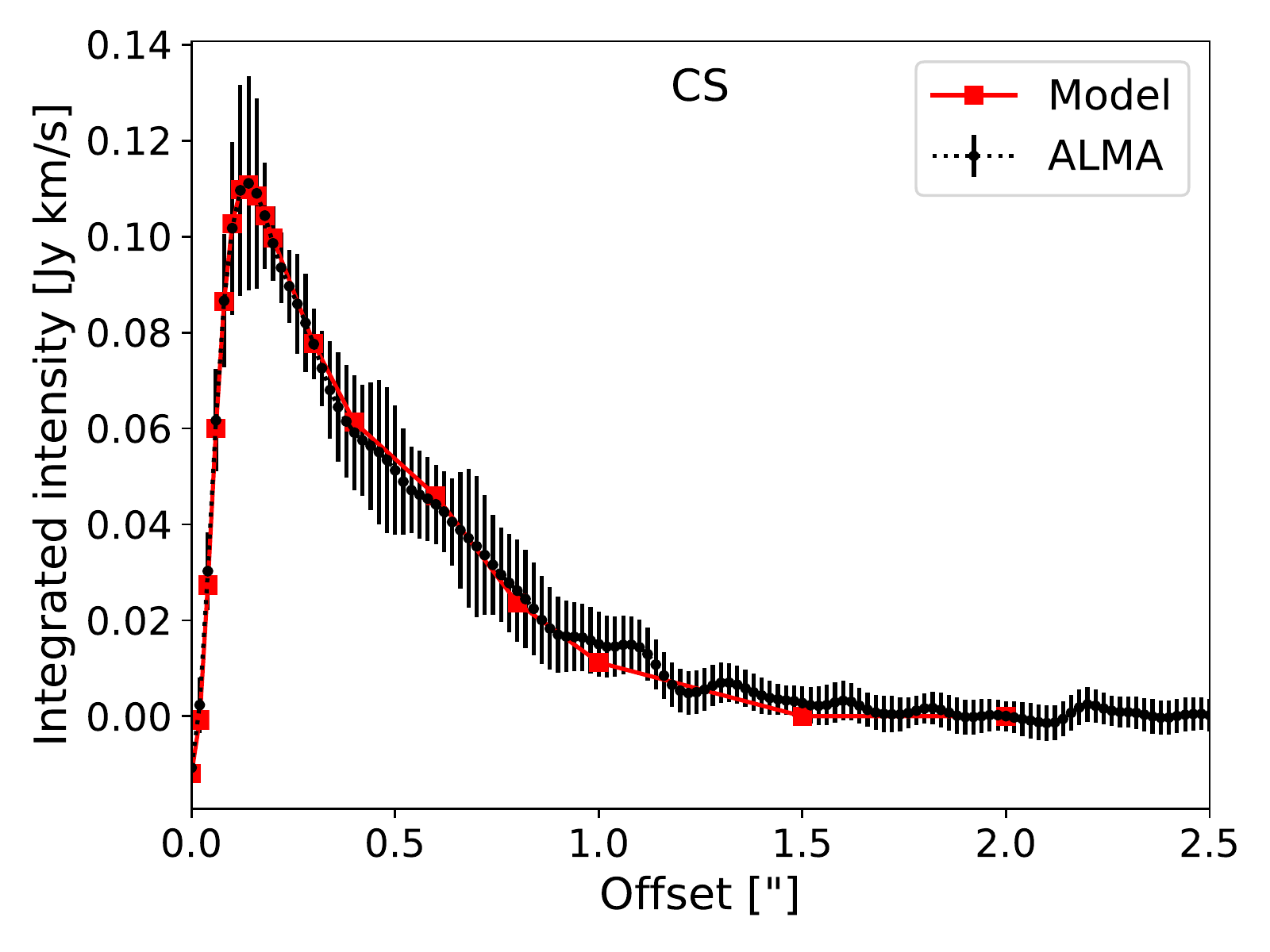}
\includegraphics[width=0.5\textwidth]{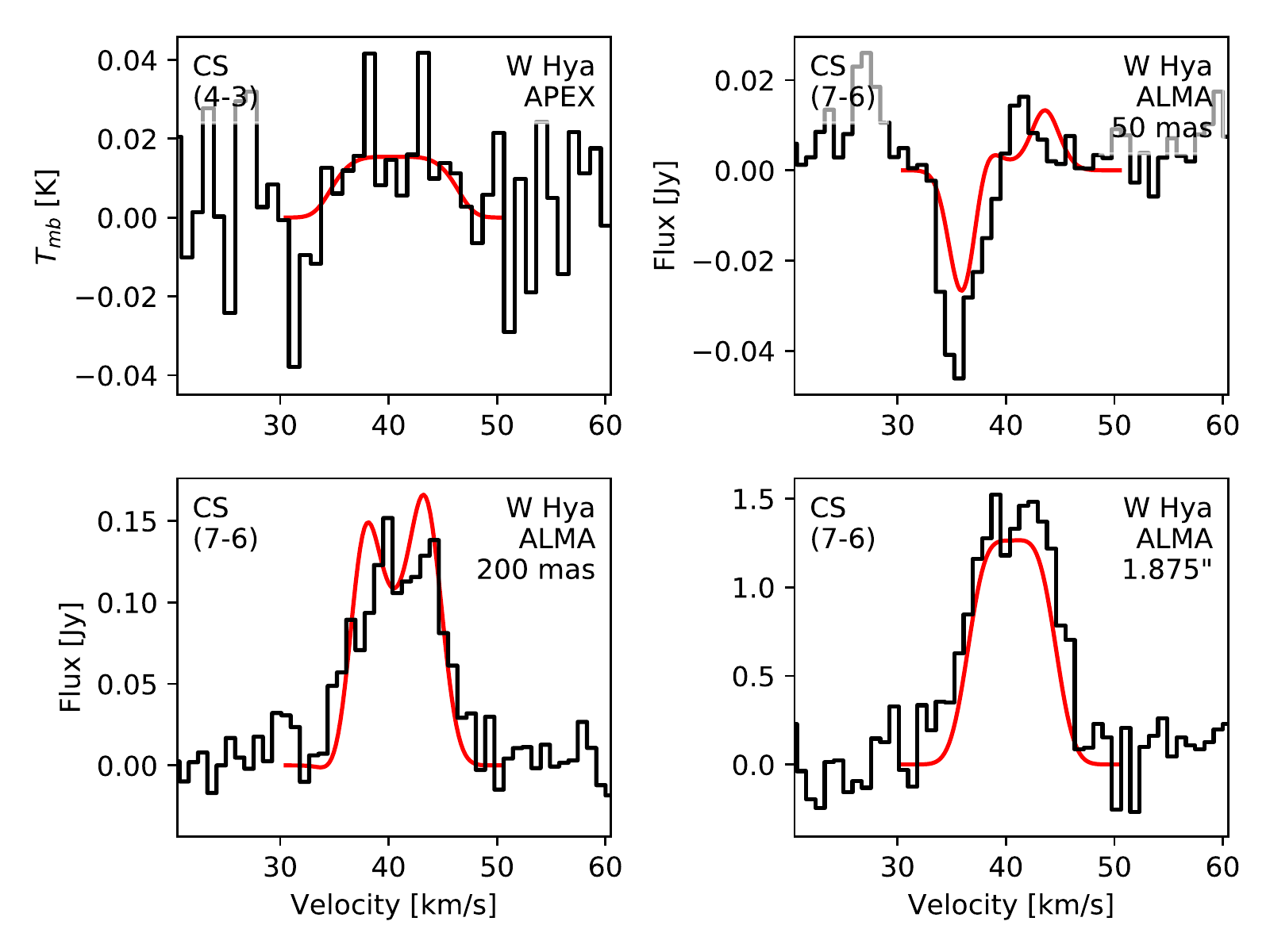}
\caption{\textit{Top:} The best CS model radial profile (red squares and solid line) plotted with the ALMA azimuthally averaged radial profile (black points and dotted line) of the ($7\to6$) transition for W~Hya. The error bars shown on the ALMA data are as described in Sect. \ref{datared}. \textit{Bottom:} The observed spectra (black histograms) and model line profiles (red lines) for CS towards W~Hya, including an undetected observation from APEX. The sizes of the extraction radii for the ALMA spectra are listed in the top right corners.\label{whyacsrad}}
\end{figure}

For W~Hya, we used both the azimuthally averaged radial profile and the CS spectra extracted from the ALMA data for different angular sizes. We also included the upper limit for the CS ($4\to3$) line from APEX observations performed by \cite{Danilovich2018}, although this did not directly contribute to constraining the model. The radial profile of the best model is plotted against the ALMA radial profile of the CS ($7\to6$) transition in Fig. \ref{whyacsrad}, along with the spectral lines. This best-fitting model is a step function from close to the surface of the star ($\sim R_*$) with an inner abundance of $f_{c_1}=4\times 10^{-9}$, an increase to $f_{c_2}=2.9\times 10^{-8}$ at $R_1=1.5\times 10^{14}$~cm and another increase to $f_g=6.5\times 10^{-8}$ at $R_2=6.5\times 10^{14}$~cm followed by a Gaussian decrease with an $e$-folding radius of $R_e=6\times 10^{15}$~cm. We found that without decreasing the inner radius of the abundance profile to the stellar radius, it was not possible to properly model the strong central absorption features seen in both the radial profile and in the 100~mas diameter spectrum in Fig. \ref{whyacsrad}. The abundance profile of CS for W~Hya is plotted in Fig. \ref{whyaabundance} and compared with the previous upper limit derived by \cite{Danilovich2018}.

\subsection{R Dor model results}

\subsubsection{SiS}

Since we do not have a radial profile for SiS towards R~Dor, due to the very faint emission of this transition, we are unable to precisely constrain the size of the SiS emitting envelope. 
The $e$-folding radius for a Gaussian distribution predicted by the SiS formula calculated in \cite{Danilovich2018} gives $R_e = 3.6\times 10^{14}$~cm which is only about six stellar radii ($6R_*$). This formula is calculated from only high mass-loss rate sources and may not hold for low mass-loss rates. Running a model with this $R_e$ did not fit the data well, even when we tried adjusting the inner radius.

Adjusting the $R_e$ to find a fit to our observations, we found that we could not properly constrain the model for $R_e$ larger than $1\times10^{15}$~cm. This is mostly due to the reduced signal-to-noise of the observations when extracting spectra for apertures with radii larger than 1\arcsec. Hence we use $R_e = 10^{15}$~cm. We also tested different inner radii: $R_\mathrm{in}=R_*$ in analogy with the W Hya results, $R_\mathrm{in}= 6.6\times10^{13}$~cm from the modelling of \cite{Maercker2016} and $R_\mathrm{in}= 1.9\times10^{14}$~cm from the modelling of \cite{Danilovich2016}. Our model results significantly over-predicted the inner emission for the two smaller radii, so we use $R_\mathrm{in}= 1.9\times10^{14}$~cm in our final model.

\begin{figure}
\includegraphics[width=0.5\textwidth]{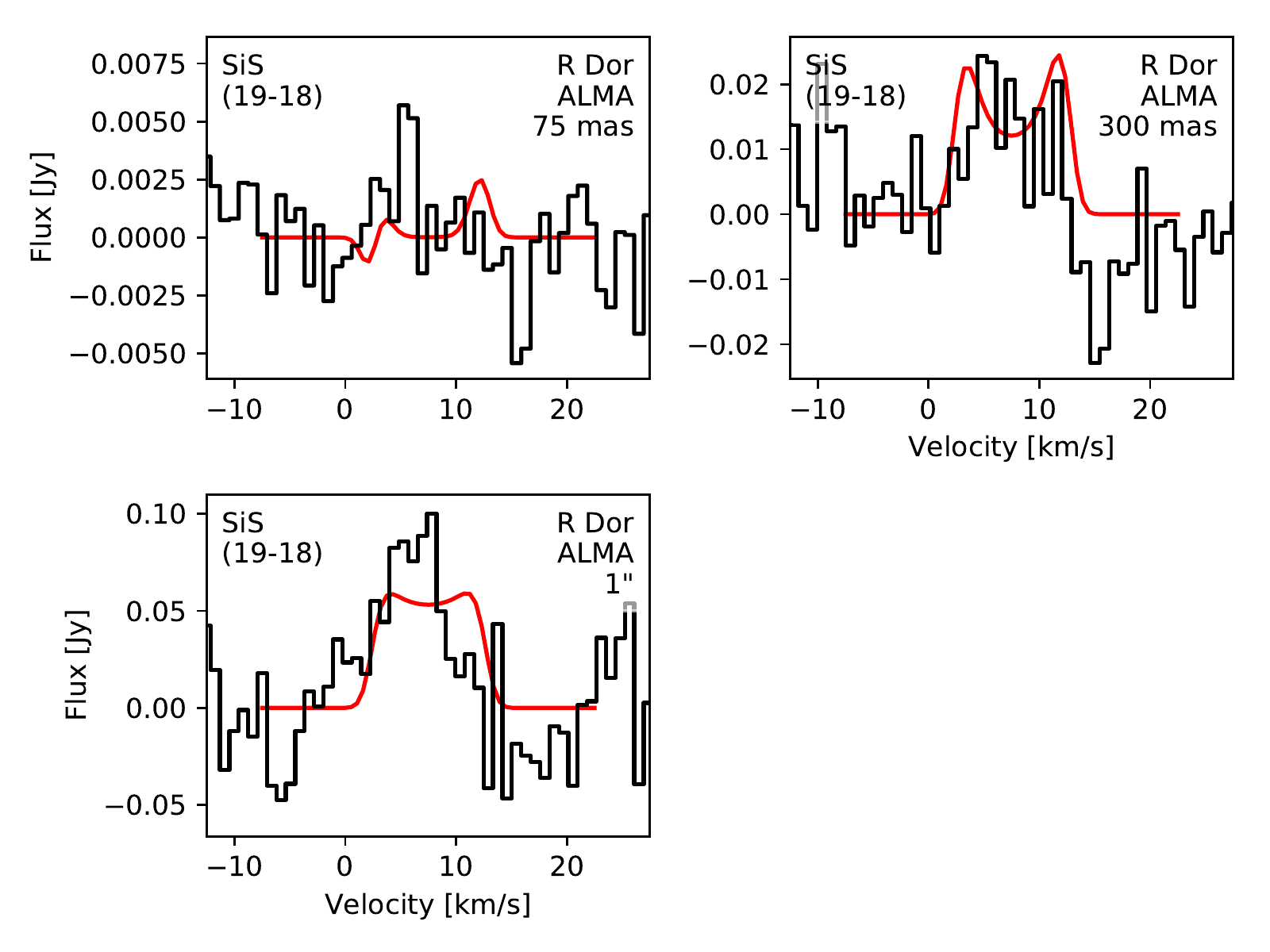}
\caption{R~Dor SiS model (red lines) and observed spectrum (black histograms). The sizes of the extraction radii for the ALMA spectra are listed in the top right corners.}
\label{rdorsis}
\end{figure}

\begin{figure}
\includegraphics[width=0.5\textwidth]{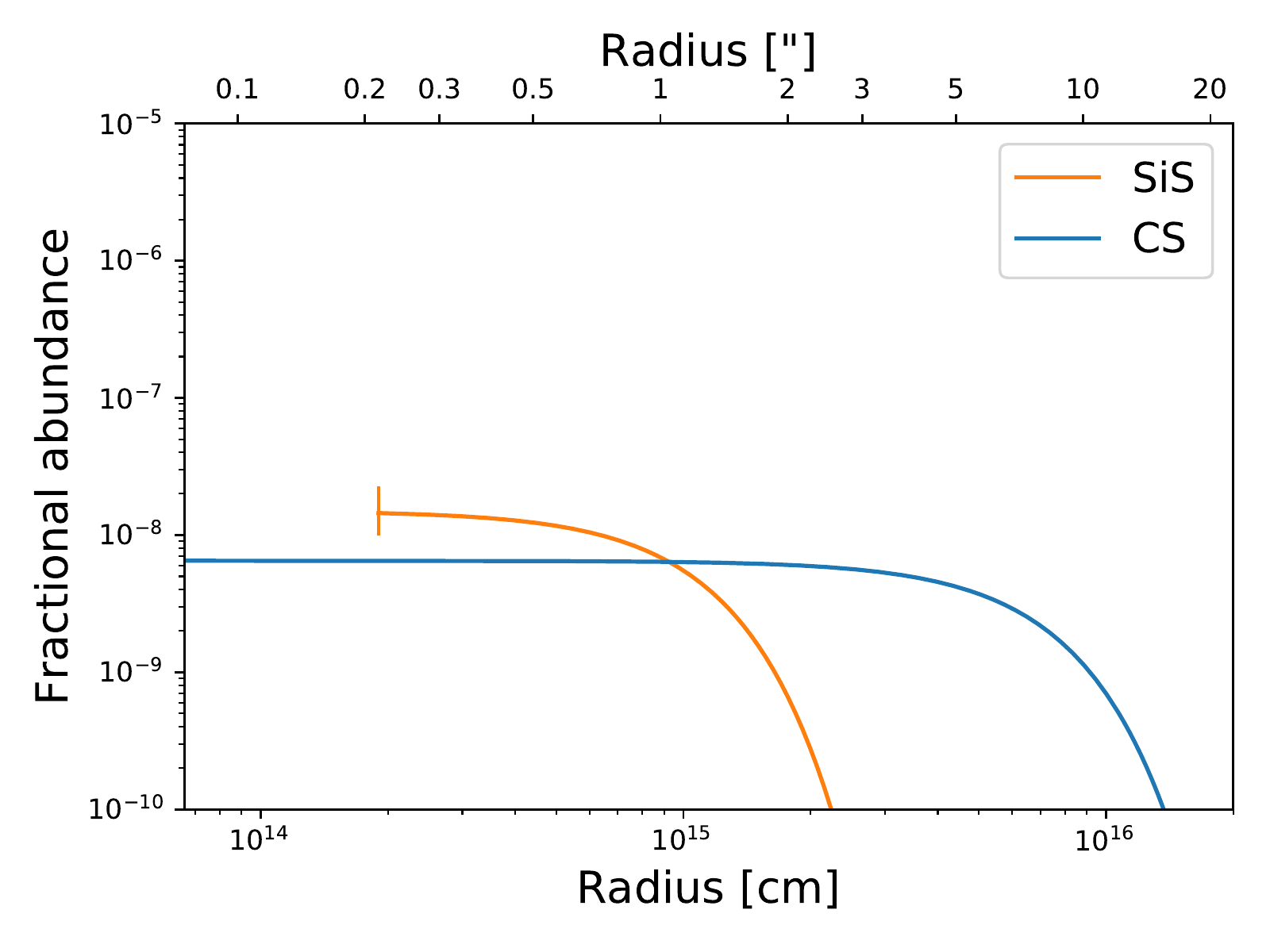}
\caption{The abundance distributions derived from modelling the R~Dor observations of CS and SiS.}
\label{rdorabundance}
\end{figure}

In all model test cases we vary the peak central abundance, $f_0$, in increments of $0.5\times 10^{-8}$ until the resultant emission lines best match the observed emission lines. For the best model we found $f_0 = 1.5\times10^{-8}$ relative to H$_2$.
 The resultant models are plotted with the observed spectra in Fig.~\ref{rdorsis} and the radial abundance distribution of the model is plotted in Fig. \ref{rdorabundance}.

\subsubsection{CS}

\begin{figure}
\begin{center}
\includegraphics[width=0.5\textwidth]{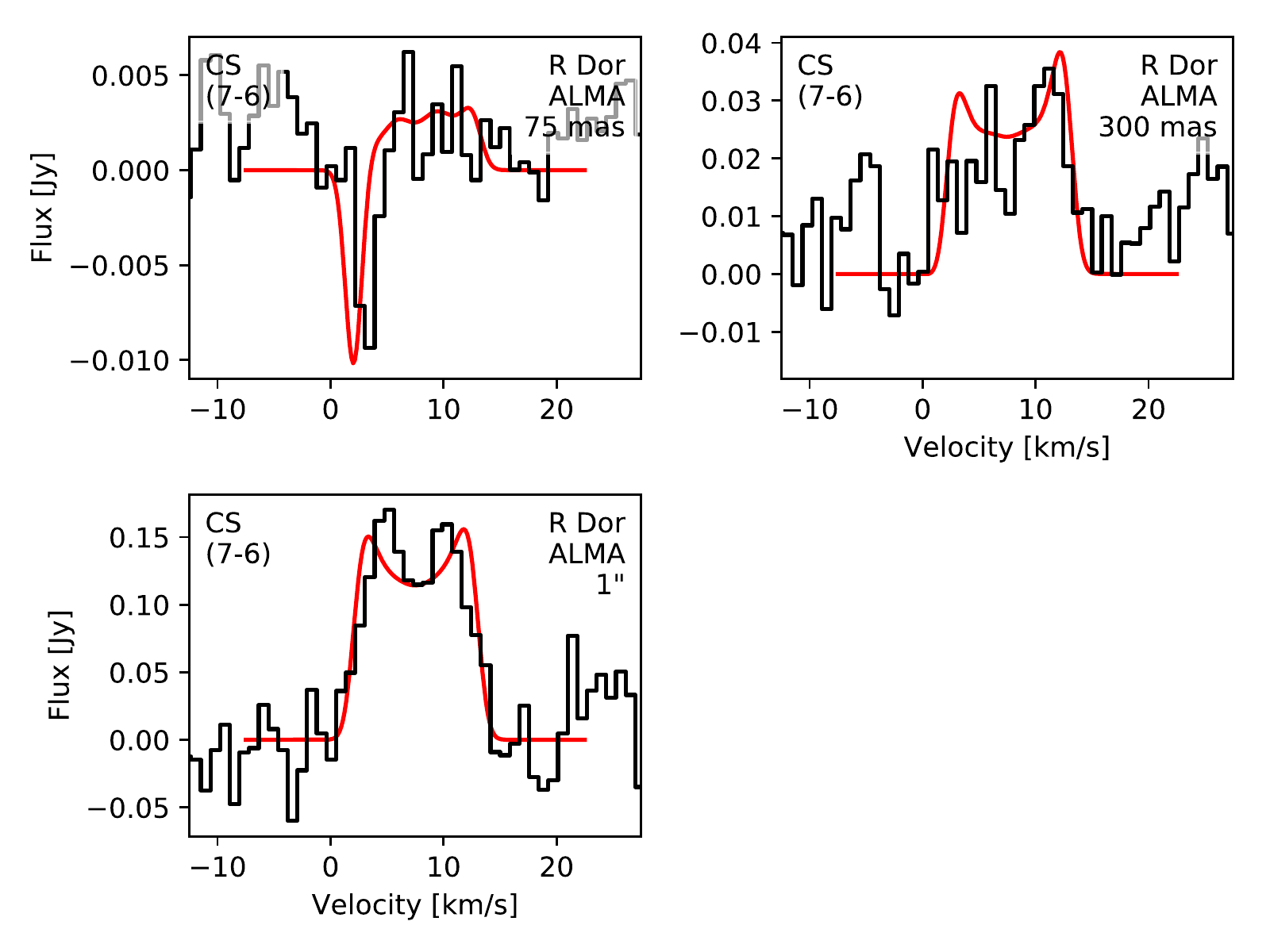}
\caption{R~Dor CS model (red lines) and observed spectrum (black histograms). The sizes of the extraction radii for the ALMA spectra are listed in the top right corners.}
\label{rdorcs}
\end{center}
\end{figure}

For R~Dor we use a similar procedure for CS as for SiS, basing our model on the ALMA spectral lines since the signal-to-noise of the observation is too low to extract a radial profile.
 We assumed a Gaussian distribution with the $e$-folding radius predicted by the CS formula calculated in \cite{Danilovich2018}. This gave $R_e = 6.7\times 10^{15}$~cm and, due to the limitations of the data, we cannot constrain the $e$-folding radius further. As for SiS, we tested three inner radii: $R_\mathrm{in}=R_*$, $6.6\times10^{13}$~cm, and $1.9\times10^{14}$~cm. Although our observations are not sensitive enough to constrain the inner radius precisely, we found that of the previous options the best fit came from $R_\mathrm{in}=6.6\times10^{13}$~cm.
By varying the peak central abundance, $f_0$, in increments of $0.5\times10^{-9}$ until the resultant emission line does not exceed the observed spectrum, we find a best fitting abundance of $f_0=6.5\times 10^{-9}$ relative to H$_2$. The resultant model is plotted with the observed spectrum in Fig. \ref{rdorcs} and the radial abundance distribution of the model is plotted in Fig. \ref{rdorabundance}. As can be seen in Fig. \ref{rdorcs}, although our model reproduces the general shape of the lines well, there is an offset of $\sim 1.5$~\kms{} between the observed absorption feature and the modelled absorption feature in the smallest extraction region spectrum. This is not due to an inaccurate LSR velocity since the spectral line from the largest extraction region is well-aligned with the model line. 
It can be accounted for if we consider the rotating disc around R~Dor proposed by \cite{Homan2018}. In a spherical expanding envelope, which is perforce what our 1D code models, we expect the absorption feature to be located between, $\upsilon_\mathrm{LSR}-\upsilon_\mathrm{min}$, where $\upsilon_\mathrm{min}$ is the minimum expansion velocity (generally taken to be the speed of sound close to the star, $\sim2$--3~\kms), and $-\upsilon_\mathrm{exp}$. This is indeed what we see in the case of W~Hya, where the modelled location of the absorption feature agrees well with the observation. In the case of a rotating disc around R~Dor, however, we expect some of the gas in the direct line of sight to the star to have a radial velocity of zero with respect to the LSR velocity, since it is instead moving laterally, assuming that the disc is edge-on or close to edge-on. Hence, the absorption feature would be found closer to $\upsilon_\mathrm{LSR}$ as is, indeed, seen in our observations. Similar arguments can be used in the case of stellar rotation causing a velocity field in the regions close to the stellar surface as suggested by \cite{Vlemmings2018} and our observations do not rule out either scenario.

\section{Discussion}

\subsection{Difference between lower and higher mass-loss rate stars}

\begin{figure*}
\begin{center}
\includegraphics[width=0.49\textwidth]{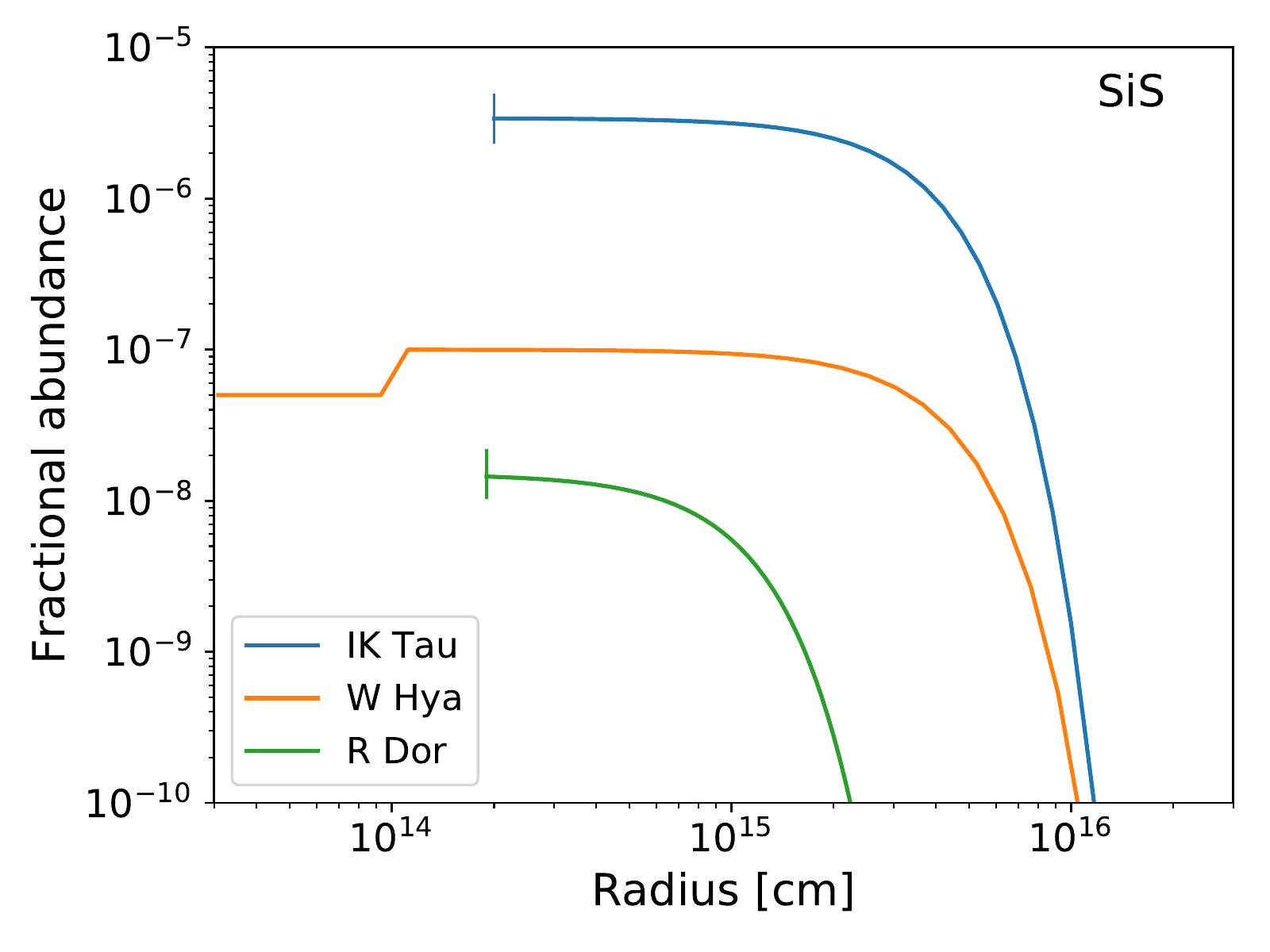}
\includegraphics[width=0.49\textwidth]{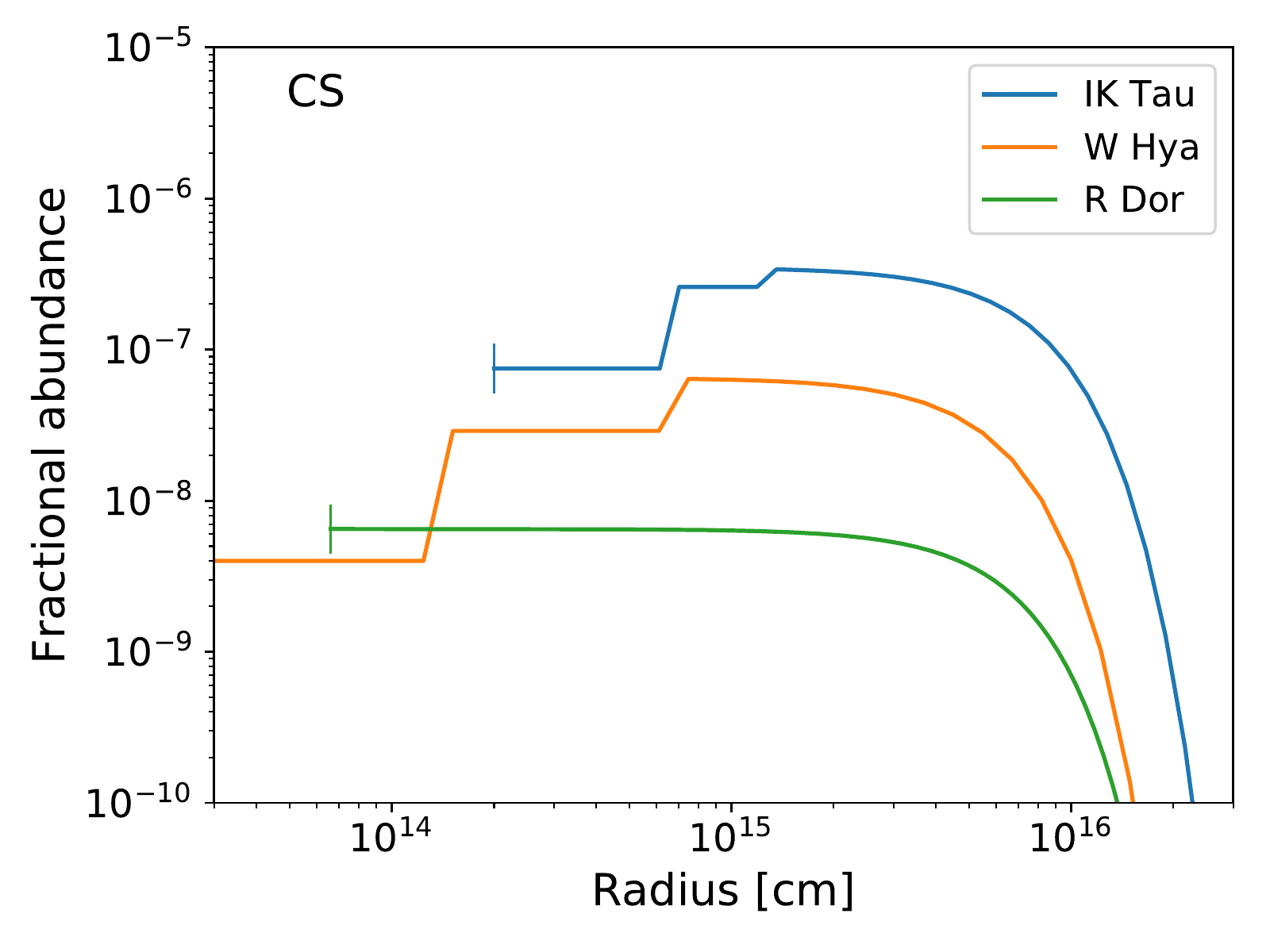}
\caption{SiS abundances (\textit{left}) and CS abundances (\textit{right}) for all three modelled stars. The vertical lines are the inner radii of the corresponding models.}
\label{allabundances}
\end{center}
\end{figure*}

In Fig. \ref{allabundances} we have plotted the radial abundance distributions derived for all three stars in our sample, grouped by molecule. For both SiS and CS, we see a progression in (peak) abundance from R~Dor to W~Hya to IK~Tau. This is unexpected if we assume that the abundances of either of these molecules is solely dependent on the mass-loss rate ($\dot{M}$) or wind density ($\dot{M}/\upsilon_\mathrm{exp}$); while IK~Tau has the highest wind density, around an order of magnitude higher than that of R~Dor, W~Hya's wind density is about a factor of two smaller than R~Dor's. This strongly implies that another factor is at play, even if, as suggested by the results of \cite{Danilovich2018}, density may contribute to the presence of SiS (across all chemical types of AGB stars) and CS (for oxygen-rich AGB CSEs). In this section we discuss some of the factors that may be at play, including density, elemental abundances, clumping and pulsation types.

SiS is thought to form in denser environments \citep{Danilovich2018}, although in general SiS chemistry is currently not very well characterised. We note that SiS was not detected towards any of the semi-regular variables included in the study of \cite{Danilovich2018}, including two of the carbon stars (U~Hya and X~TrA) the rest of which generally had higher SiS abundances than the oxygen-rich stars. There is ample evidence that significantly lower SiS abundances are found for both lower mass-loss rates and less extreme pulsators. For example, \cite{Schoier2007} also did not detect SiS in any semi-regular stars except for RW LMi, a SRa with a long period (640 days) and a V-band amplitude of 3.7 mag, which is in the Mira variable range. RW~LMi is also the only SR star with an SiS detection in the survey of \cite{Bujarrabal1994}; the rest are all Miras. This suggests that SiS forms more readily and reaches higher abundances in the environments of Mira variable CSEs rather than in the CSEs of semi-regular variables. Since semi-regular variables are characterised by smaller amplitudes ($\lesssim 2.5$~mag in the V band), while Miras are characterised by larger amplitudes ($\gtrsim 2.5$~mag) and regular periods, the amplitude could have an impact on SiS formation. The tentative correlation between SiS abundance and pulsation amplitude that we see in this present study (limited by the small sample) could be due to shocks, as with CS, but earlier studies suggest that it may hold across chemical types (the study of \cite{Schoier2007} included carbon stars and those of \cite{Danilovich2018} and \cite{Bujarrabal1994} included carbon and S-type stars), while the shock-induced formation of CS is most relevant to oxygen-rich stars.% since CS can form more easier in carbon-rich CSEs.

We also consider the sulphur and silicon budgets, since these two elements are not nucleosynthesised in AGB stars nor their main sequence progenitors. Assuming a solar abundance of both elements \citep{Asplund2009}, past studies have shown that the sulphur budget of R~Dor can be almost entirely accounted for by SO and SO$_2$ \citep{Danilovich2016}, while the silicon budget is similarly accounted for by SiO \citep{Van-de-Sande2018a}, not leaving much scope for the formation of SiS (note that the carbon budget is more complicated and cannot readily be estimated from the solar abundance of carbon). The same results are found for W~Hya by \cite{Khouri2014a} for Si and \cite{Danilovich2016} for S, which is consistent with our results here; the amount of Si and S accounted for by our SiS and CS models is negligible compared with the SiO, SO, and SO$_2$ abundances {in these low mass-loss rate stars}. IK~Tau differs from the other two stars in many ways. The SO and SO$_2$ abundances found by \cite{Danilovich2016} are roughly an order of magnitude lower than for R~Dor and W~Hya and are comparable to the SiS abundance we find here. The SiO abundance found by \cite{Decin2010} is also around an order of magnitude lower than the SiO abundances of R~Dor and W~Hya and, while this could account for the increased SiS abundance \citep[as discussed by][]{Danilovich2018}, there remains plenty of silicon outside of these two molecules --- possibly having condensed into dust. 

%Marie's comment:
%----------------
%Binding energy onto dust
%
%CS = 1900 K
%SiS = 3800 K
%
%for reference, H2O = 4800 K so SiS could stick to dust quite a bit. I could run a model

It is also possible that a clumpy circumstellar medium accounts for some of the differences we see between the stars in our sample. For example, \cite{Van-de-Sande2018b} show that clumpy outflows can produce a significant amount of CS. Various channel maps of other molecules towards R~Dor show arc-like structures in the CSE \citep{Decin2018} rather than a smooth outflow, indicating a level of clumpiness in the outflow. For the CS emission in IK~Tau the position of the abundance increase can be explained by a clumpy outflow, as shown by \cite{Van-de-Sande2018b}, whereas the smooth outflow models of \cite{Willacy1997} give a CS increase further out in the envelope (between $\sim1\times10^{16}$ and $\sim3\times10^{16}$~cm) which does not agree with our results. 

Another source of differences between the stars could be the pulsation types.
Consider the fact that R~Dor is a SRb variable with at least two pulsation modes \citep[][]{Bedding1998a}, while W~Hya and IK~Tau are Mira variables \citep{Pojmanski2002,Wozniak2004}. Some variability properties of these stars, taken from the aforementioned citations and the VSX database\footnote{International Variable Star Index database, \url{www.aavso.org/vsx/}} are given in Table \ref{variability}, including period and variations in V magnitude ($\Delta V$). From this we can see that both period and amplitude increase with the progression from R~Dor to W~Hya to IK~Tau. 
CS is not a species predicted to form in oxygen-rich environments when only equilibrium chemistry is considered \citep{Cherchneff2006}, whereas it is expected to be form easily in carbon-rich environments. In oxygen-rich CSEs it is thought that shocks play a significant role in CS formation \citep[see ][for further discussion]{Cherchneff2006,Gobrecht2016,Danilovich2018}, hence it follows logically that more CS would form in stars with more extreme (larger amplitude) pulsations. %This is especially so for CS formed in the innermost regions of the star (such as W~Hya) since it is in these regions that shocks are strongest. 

Aside from the differences in abundances, our results also show differences in the inner radii of the models. Our IK~Tau models have a larger inner radius ($R_\mathrm{in}=2\times 10^{14}$~cm), while the W~Hya models start close to the stellar surface. Thanks to the high spatial resolution of the ALMA data, we are reasonably confident that this is a real difference between the stars (though we are less certain about the R~Dor inner radii results). We tested models with smaller inner radii for IK~Tau and, although little difference is apparent in the models for the APEX data, there is a noticeable difference in the models for the resolved ALMA lines and in the radial profile. If anything, the ALMA data points towards a slightly larger inner radius ($R_\mathrm{in}\approx2.2\times 10^{14}$~cm). This was not the case for W~Hya, for which we found the best agreement with an abundance profile that starts from the innermost regions close to the star. This dichotomy could point to the effects of dust formation in these CSEs. If more silicate dust is formed in IK~Tau, the production of SiS at a larger radius in IK~Tau than in W~Hya could be due to sputtering dust grains making Si available for SiS formation. This theory is supported by the inner radius of our IK~Tau SiS model lying close to where dust has formed in the CSE \citep{Decin2017}, and by the fact that using a smaller inner radius for our model did not agree with the observations.

%SiS has been often considered as a parent species \citep[one formed closer to the star than the inner radius of the chemical model used, e.g. by][]{Willacy1997,Li2016} but our results suggest that it is not in the case of IK~Tau, but seems to be for W~Hya. For CS, a possible cause of the difference between IK~Tau and W~Hya could be the higher optical depth of IK~Tau's CSE leading to fewer UV photons penetrating to the inner regions and hence hampering CS formation through UV photochemistry \citep[as indicated by the chemical modelling of][]{Van-de-Sande2018b}. 

The question that remains is: why are such different proportions of the sulphur molecules present in these oxygen-rich stars? Is it a case of evolutionary change --- are the differences due to the differing ages of the stars? If that is the case, then what drives the change over time? Is it due to dust formation or dust-gas chemistry or other physical changes in the CSE? 
%For example, \cite{Decin2017} plot the infrared spectra of R~Dor and W~Hya together, showing a significant variation in dust emission around $\sim$10--20~\micron. 
Or is another factor at play?

\begin{table}
\caption{Variability properties of studied stars}\label{variability}
\begin{center}
\begin{tabular}{cccc}
\hline\hline
 Star & $\Delta V$ [mag] & Period(s) [days] & Type\\
\hline
IK Tau & 5.7 & 470 & Mira\\
W Hya & 4.0 & 390 & Mira\\
R Dor & 1.54 & 172, 344 & SRb\\
\hline
\end{tabular}
\end{center}
Values taken from the VSX database.
\end{table}

\subsection{Comparison with previous studies}

\cite{Danilovich2018} observed CS and SiS lines using APEX and IRAM for a large sample of AGB stars, including IK~Tau and W~Hya, the latter of which did not yield any detected lines. Based on these observations and radiative transfer modelling, they derived empirical formulae for the $e$-folding radii of both molecules based on mass-loss rate and terminal expansion velocity. Using these formulae, they also found upper limits for W~Hya based on their APEX non-detections, which we plot with our ALMA results in Fig. \ref{whyaabundance}. As can be seen there, both upper limits assumed inner radii of $2\times10^{14}$~cm and both gave higher fractional abundances than we derive from the ALMA data. For CS, the $e$-folding radius predicted by the formula is in good agreement with the radius we found in the Gaussian part of our model ($5.6\times10^{15}$~cm compared with our $6\times10^{15}$~cm) and the peak fractional abundance was only a little higher than the peak abundance for the Gaussian part of our model. The inner regions of our model stretch to the stellar surface and were found to be significantly lower than the upper limit model. There is a more significant difference between their SiS upper limit and our model, however. The $e$-folding radius predicted by their formula is an order of magnitude smaller than what we found based on the ALMA data and our peak fractional abundance is an order of magnitude lower than the upper limit they found based on the APEX observations. This suggests that the SiS $e$-folding radius formula found by \cite{Danilovich2018} does not hold for low mass-loss rate stars.

In the case of IK~Tau, the model calculated by \cite{Danilovich2018} for SiS was in agreement with our more precise ALMA results to within a factor of $\sim 2$ for both peak fractional abundance and $e$-folding radius. For CS we found a less regular distribution of CS than was apparent from the single dish observations and hence our model deviates from that found by \cite{Danilovich2018} more significantly: the Gaussian portion of our model has an $e$-folding radius a factor of $\sim 4$ smaller and our peak fractional abundance is a factor of $\sim 3$ larger.

IK~Tau has been the subject of other single-dish studies of these two molecules. \cite{Schoier2007} surveyed SiS in a large sample carbon- and oxygen-rich AGB stars and modelled the SiS emission. They assumed $e$-folding radii for SiS based on the empirical result derived by \cite{Gonzalez-Delgado2003} for SiO, giving an $e$-folding radius for IK~Tau of $1.6\times10^{16}$~cm, almost three times larger than ours. To find adequate fits to their observations, they also implemented a compact central component of SiS out to $1\times10^{15}$~cm with a high SiS abundance of $2\times10^{-5}$ in this region. We do not see such a compact inner region in our ALMA observations, which are certainly sensitive enough to detect the kind of jump in abundance they used. 

\cite{Decin2010} modelled several different molecular species towards IK~Tau, including SiS and CS. They implemented a similar SiS model to \cite{Schoier2007}, with a more abundant inner component and a lower abundance outer component, the latter being much larger than the size of our SiS radial abundance distribution. Their inner abundance is also larger than ours, exceeding it by more than a factor of three. Their CS model gives a different radial abundance profile to ours, with a constant inner abundance paired with a rise and decline in the outer regions. Their fractional abundance is much smaller than what we found in the outer regions of our model, although it agrees well with the innermost abundance we found. \cite{Decin2010} also calculated isotopic ratios for Si from SiO observations. Their results agree with ours within the specified errors and they included a literature review and discussion of Si isotopic ratios in their Sect. 5.2, to which we direct interested readers.
%28 SiO/29 SiO = 27, 28 SiO/30 SiO = 80, and 29 SiO/30 SiO = 3, with an uncertainty of a factor of ?2 
% Remember that Decin 2010 gives abundances relative to H not H2

\cite{Velilla-Prieto2017} performed a line survey of IK~Tau using the IRAM 30~m telescope and detected $\sim350$ lines across 34 molecular species. They also calculated Si and S isotopic ratios based on their observations of SiO, SiS, CS, SO, and SO$_2$ and on abundances they derived from population diagrams. The abundances they found for $^{28}$Si$^{32}$S and $^{12}$C$^{32}$S are within 30\% and 50\%, respectively, of our results. However their abundances of the less common isotopologues of these molecules are factors of 3--4 higher than our results. Hence their isotopologue ratios are not in agreement with ours. 
They find $^{32}$S/$^{34}$S = 12.5 from CS and SiS (and lower values from SO and SO$_2$), which is significantly lower than our value of $37\pm14$. For $^{28}$Si/$^{29}$Si calculated from SiS, they find 11, which is also much smaller than our value of $25\pm10$. For $^{28}$Si/$^{30}$Si, also calculated from SiS, they find 16, less than half of our value of $40\pm8$. 
%These differences arise from differences in their abundances for the rarer isotopologues compared with our results here. While their $^{28}$Si$^{32}$S abundance is similar to ours (less than 30\% larger), their $^{29}$Si$^{32}$S and $^{30}$Si$^{32}$S abundances are more than three times higher and their $^{28}$Si$^{34}$S abundance is more than four times higher than ours. Similar trends are seen for CS with the \cite{Velilla-Prieto2017} abundances exceeding ours most significantly for the rarer isotopologue.

\cite{Peng2013} observed the less abundant isotopologues of SiO and hence calculated the $^{29}$Si/$^{30}$Si ratio for a sample of stars, including IK~Tau. They found a ratio of $1.60 \pm 0.30$, in good agreement with our result of $1.7\pm0.3$.

\cite{Brunner2018} modelled several molecules observed by ALMA towards the S-type AGB star W~Aql, including CS, SiS and $^{30}$SiS. Their observations are of lower spatial resolution than ours and no deviations from Gaussian abundance profiles can be seen in the inner regions of their observations. They do find the need to include an overdensity in their models at around $10^{16}$~cm from the centre of the star, but this is assumed to be due to spiral windings seen more clearly in the CO ALMA observations of W~Aql presented by \cite{Ramstedt2017}. Overdensity aside, their Gaussian SiS abundance profile is similar to what we found for IK~Tau --- suggesting that the molecule may behave similarly in the circumstellar envelopes of these chemically different stars with similar mass-loss rates. They found a higher abundance of CS, however, which is not surprising given that W~Aql is an S-type star with a higher abundance of C than is present around IK~Tau, an oxygen-rich star. They found a much lower $^{28}$SiS/$^{30}$SiS ratio of 13 for W~Aql than we found for IK~Tau (40) and a much lower $^{28}$Si/$^{29}$Si ratio of 11.6 (from SiO) compared with our ratio of 25. These ratios are indicative of W~Aql having a higher metallicity than IK~Tau does \citep[see, for example, the model results of][]{Kobayashi2011}. The $^{29}$Si/$^{30}$Si ratio for W~Aql, derived by \cite{Brunner2018} by combining SiO and SiS observations, is also smaller than what we find for IK~Tau, with 1.1 for W~Aql compared with 1.7 for IK~Tau. Since we do not expect the $^{29}$Si/$^{30}$Si ratio to change much with metallicity \citep{Kobayashi2011}, this could be a result of inhomogeneous chemical evolution of the galaxy rather than a tracer of metallicity.
%\todo[inline,color=green!20]{Or does it indicate something else?}
%Amanda says: 
%From looking at Fig 17 from Kobayashi et al. (2011) who show the evolution of the Si isotopes, it looks like Si29 and Si30 increase in lock-step together, when the nucleosynthesis is dominated by massive stars. 
%
%The s-process in AGB stars will change the Si isotopic ratio of Si29/Si30. The ratio is lowered but in my models not greatly so, from ~1.6 to 1.4 at most. 
%
%So I'm not sure I would intrepet the above variation in si29/si30 as caused by variations in metallicity. It could also be caused by stochastic/inhomogenous chem evolution in the Galactic disc. 

\subsection{Isotopologue ratios and galactic chemical evolution}

In Table \ref{isotopes}, we compare our calculated IK~Tau isotopic ratios with literature values for R~Dor and W~Hya and the solar isotope ratios from \cite{Asplund2009}. Our IK~Tau values are generally larger than the solar values and the R~Dor values from the literature \citep{Danilovich2016,De-Beck2018}. Since we have used the convention of always writing the ratios as lower mass number over higher mass number, this means that we found IK~Tau to have a lower proportion of heavier isotopes than the Sun and R~Dor. This difference is most noticeable when comparing rarer isotopologues to the most common isotopologue, while the ratios between the less common isotopes are in good agreement with the solar values. For silicon, our $^{28}$Si/$^{29}$Si and $^{28}$Si/$^{30}$Si ratios are close to 30\% higher than the solar value, and for sulphur our $^{32}$S/$^{33}$S ratio is about 50\% higher than the solar value and the $^{32}$S/$^{34}$S ratio is almost 70\% higher. Our $^{29}$Si/$^{30}$Si ratio is only 13\% higher than the solar value while $^{33}$S/$^{34}$S ratio is within 5\% of the solar value. The general trend suggests that IK~Tau has a lower metallicity than the Sun \citep[the same conclusion was reached by][]{Decin2010}. Conversely, the literature data suggests R~Dor may have a slightly higher metallicity than the Sun when considering the Si isotopes, although the difference is less pronounced (and the $^{32}$S/$^{34}$S ratio is in good agreement with the solar value).

$^{28}$Si, $^{32}$S, and $^{34}$S are produced through oxygen burning and in Type II supernovae (SNe II) through explosive nucleosynthesis \citep{Anders1989,Hughes2008}. The less abundant isotopes of $^{29}$Si and $^{30}$Si are formed through neon burning and neutron capture in SNe II \citep{Timmes1996,Zinner2006} and may increase in abundance during the AGB phase through slow neutron captures (the $s$-process). 
$^{34}$S is partly also formed from neutron captures via $^{33}$S, which means its abundance (and the abundance of $^{33}$S) may increase in AGB stars \citep{Hughes2008}. 
%With regard to galactic chemical evolution, \cite{Kobayashi2011} show that 
However, in the models of \cite{Karakas2016}, the $^{32}$S/$^{33}$S and $^{32}$S/$^{34}$S ratios do not change much over the course of AGB evolution, suggesting that these can be readily used to trace galactic chemical evolution. For the most extreme example of a low-mass AGB star, a star with a main sequence mass 3~\msol and metallicity $Z = 0.014$ that becomes carbon-rich and experiences considerable dredge up will only have a shift of $\sim6$\% in $^{32}$S/$^{33}$S and a shift of $\sim1.3$\% in $^{32}$S/$^{34}$S while on the thermally pulsing AGB.
 \cite{Karakas2016} predict that the $^{32}$S/$^{36}$S ratio does change more significantly due to neutron captures in the He-shell. We did not detect Si$^{36}$S in the ALMA scan and C$^{36}$S falls outside of the observed frequency range. (Note also that none of the strongest $^{36}$SO lines fall within the observed frequency range.)
A quick analysis of Si$^{36}$S based on our non-detections with ALMA gives an upper-limit abundance $f_0<10^{-9}$ relative to H$_2$. This gives a lower-limit $^{32}$S/$^{36}$S that is about 30\% lower than the solar value.

\cite{Chin1996} find a gradient in $^{32}$S/$^{34}$S ratio with distance from the galactic centre (though not for $^{34}$S/$^{33}$S). However, their observations are of star-forming regions (with $^{32}$S/$^{34}$S ranging from $\sim$14 to 35) and the solar system value does not lie very close to their trend line. Since star-forming regions are in a very different evolutionary phase to AGB stars (and to main-sequence stars like the Sun), it is unclear that their isotopologue ratios can be directly related to those that we find for individual stars. Aside from the spread in the age-metallicity relationship in the solar neighbourhood \citep{Feltzing2001}, IK~Tau (and the Sun) formed from molecular clouds that significantly predate present-day star-forming regions and hence are not comparable. Overall, our results are in agreement with the conclusions drawn by \cite{Decin2010} that, based on isotopic ratios, the ISM from which IK~Tau formed was enriched by SNe II \citep[see also][]{Zinner2006}.

\section{Conclusions}

We analysed ALMA observations of SiS and CS emission lines for IK~Tau, W~Hya, and R~Dor and were successfully able to use radiative transfer modelling to derive abundance distributions for all three. We found that of the three stars R~Dor had the lowest abundances for both molecules and IK~Tau had the highest, with the difference in peak abundance between the two extremes spanning more than two orders of magnitude for SiS and 1.5 dex for CS. For CS towards IK~Tau and both molecules towards W~Hya, we found stratified abundance distributions. The W~Hya emission and models also indicates that both of these molecules are found very close to the star, while the IK~Tau models show these molecules forming further out in the CSE of IK~Tau. 

We also calculated abundances for several isotopologues detected towards IK~Tau: C$^{34}$S, $^{29}$SiS, $^{30}$SiS, Si$^{33}$S, Si$^{34}$S, $^{29}$Si$^{34}$S, and $^{30}$Si$^{34}$S.  Overall the isotopic ratios we derived from these suggest a lower metallicity for IK~Tau than the solar value.

\section*{Acknowledgements}

LD, MVdS and FDC acknowledge support from the ERC consolidator grant 646758 AEROSOL. LD acknowledges support from the FWO Research Project grant G024112N. MVdS acknowledges support from the FWO. FDC is supported by the EPSRC iCASE studentship programme, Intel Corporation and Cray Inc.
This paper makes use of the following ALMA data: ADS/JAO.ALMA 2013.0.00166.S and 2015.1.01446.S. ALMA is a partnership of ESO (representing its member states), NSF (USA) and NINS (Japan), together with NRC (Canada) and NSC and ASIAA (Taiwan), in cooperation with the Republic of Chile. The Joint ALMA Observatory is operated by ESO, AUI/NRAO and NAOJ.
This research has made use of the International Variable Star Index (VSX) database, operated at AAVSO, Cambridge, Massachusetts, USA.

%%%%%%%%%%%%%%%%%%%%%%%%%%%%%%%%%%%%%%%%%%%%%%%%%%

%%%%%%%%%%%%%%%%%%%% REFERENCES %%%%%%%%%%%%%%%%%%

% The best way to enter references is to use BibTeX:

%\bibliographystyle{mnras}
%\bibliography{example} % if your bibtex file is called example.bib

% Alternatively you could enter them by hand, like this:
% This method is tedious and prone to error if you have lots of references
%\begin{thebibliography}{99}
%\bibitem[\protect\citeauthoryear{Author}{2012}]{Author2012}
%Author A.~N., 2013, Journal of Improbable Astronomy, 1, 1
%\bibitem[\protect\citeauthoryear{Others}{2013}]{Others2013}
%Others S., 2012, Journal of Interesting Stuff, 17, 198
%\end{thebibliography}

\bibliographystyle{mnras}
\bibliography{ALMACSSiS}
\bsp	% typesetting comment
\label{lastpage}
\end{document}